\documentclass[letterpaper,twocolumn,10pt]{article}
\usepackage{usenix}

\usepackage{tikz}
\usepackage{amsmath}
\usepackage{xspace}
\usepackage{subcaption} 
\usepackage{algorithm}
\usepackage{algpseudocode}
\usepackage{siunitx}

\newcommand{\sys}{Nixie\xspace}
\newcommand{\daemon}{\sys Daemon\xspace}
\newcommand{\shim}{\sys Shim\xspace}

\def\Snospace~{\textsection}

\begin{document}

\date{}

\title{\Large \bf \sys: Efficient, Transparent Temporal Multiplexing for Consumer GPUs}

\author{
{\rm Yechen Xu\quad Yifei Wang\quad Nathanael Ren\quad Yiran Chen\quad Danyang Zhuo}\\
Duke University
}

\maketitle


\newcommand{\NeedComparisonFigure}[2]{{\color{orange}[Figure compare with (#1): #2]}}
\newcommand{\NeedFigure}[1]{{\color{purple}[Figure: #1]}}

\newcommand{\NeedNumber}[1]{{\color{blue}[Number: #1]}}

\begin{abstract}
Consumer machines are increasingly running large ML workloads such as large language models (LLMs), text-to-image generation, and interactive image editing. Unlike datacenter GPUs, consumer GPUs serve single-user, rapidly changing workloads, and each model's working set often nearly fills the GPU memory. As a result, existing sharing mechanisms (e.g., NVIDIA Unified Virtual Memory) perform poorly due to memory thrashing and excessive use of CPU pinned memory when multiple applications are active.

We design and implement \sys, a system that enables efficient and transparent temporal multiplexing on consumer GPUs without requiring any application or driver changes. \sys is a system service that coordinates GPU memory allocation and kernel launch behavior to efficiently utilize the CPU-GPU bi-directional bandwidth and CPU pinned memory. A lightweight scheduler in \sys further improves responsiveness by automatically prioritizing latency-sensitive interactive jobs using MLFQ-inspired techniques. Our evaluations show that \sys improves latency of real interactive code-completion tasks by up to $3.8\times$ and saves up to 66.8\% CPU pinned memory usage given the same latency requirement.
\end{abstract}
\section{Introduction}

Modern consumer machines increasingly run large and sophisticated machine learning (ML) workloads. Consumers now routinely run large language models (LLMs)~\cite{qwen3, gemma3}, text-to-image diffusion models~\cite{flux1kontext}, interactive image-editing~\cite{qwenimage}, and video processing~\cite{qwen3vl, wan} locally on their desktops equipped with consumer GPUs, such as NVIDIA’s RTX 4090 and 5090. This trend is driven by the desire for privacy, interactivity, and reduced hardware costs, as well as by rapid advances in model quantization and optimized inference runtimes~\cite{llamacpp, sglang, vllm} that make local ML execution practical.

Despite this wave of local ML execution, today’s consumer GPUs are fundamentally designed for a single active application. In contrast to datacenter GPUs, where workloads are typically long-running, batched, and multi-tenant, consumer workloads are rapidly changing, heterogeneous, and strongly user-driven. A single user may frequently switch between very different models and applications: querying an LLM, generating an image, editing a photo, or running a background batch job such as OCR or video processing. 
Batch sizes for interactive applications are almost always one, and users expect low-latency responses for interactive workloads. These characteristics create a unique resource-management environment that differs substantially from datacenter GPUs.

A central challenge is that \textit{the working set of each ML application often nearly saturates GPU memory}. Models such as modern LLMs or diffusion pipelines require tens of gigabytes of active parameters, activations, and temporary workspace. User often prefer more accurate and larger models. As a result, when two or more applications run concurrently, the combined working set trivially exceeds GPU memory.

Existing GPU multiplexing mechanisms perform poorly under these conditions.
Naive stop-and-restart approaches such as llama-swap~\cite{llama-swap} introduce substantial delays: each switch requires restarting applications and reloading large models, leading to multi-second or even more response times. Application-level model swapping (e.g., Ollama~\cite{ollama}) is restricted to a single application and cannot coordinate with other applications. Prior systems such as Prism~\cite{prism} and Aegaeon~\cite{aegaeon} require manual integration and primarily target scenarios with many small models sharing a large GPU, which does not reflect the consumer GPU setting where each model’s working set nearly fills the GPU memory.

NVIDIA Unified Virtual Memory (UVM)~\cite{uvm} provides a form of transparent GPU multiplexing, but it relies on demand paging and implicitly assumes that the combined working set fits in GPU memory to sustain good performance. When multiple large applications are active, UVM exhibits severe thrashing: pages repeatedly migrate between GPU and CPU memory, causing drastic throughput degradation and large, unpredictable latency spikes. Systems such as TGS~\cite{TGS} and NVShare~\cite{nvshare} mitigate some of UVM’s thrashing behavior, but their workload assumptions remain restrictive and do not reflect consumer GPU usage patterns. Moreover, they inherit UVM’s fundamental limitations, including excessive use of CPU pinned memory, which significantly limits scalability for large, memory-intensive models.

We design and implement \sys, a system service that provides efficient and transparent temporal multiplexing on consumer GPUs without requiring any application or driver changes. \sys interposes on GPU memory allocation and kernel-launch behavior. The key idea behind \sys is to explicitly control when applications occupy GPU memory and when they yield it, ensuring that a single application's working set fully resides in GPU memory at any given moment. This avoids the pathological thrashing behavior inherent to Unified Memory and maintains predictable performance for each application. For fast context switching, \sys fully utilizes the bi-directional bandwidth between GPU and CPU.

\sys also incorporates a lightweight scheduler that automatically identifies latency-sensitive interactive applications and prioritizes them, without requiring any user annotations on application priorities. \sys uses an MLFQ-like scheduler. It dynamically adapts application priorities by tracking applications' CUDA kernel launch patterns and prioritizes applications that can finish executions within their allocated time window. Lower priority applications have larger time windows to minimize the context switch frequency and amortize scheduling overhead.

We evaluate \sys on a consumer-grade NVIDIA RTX 5090 GPU using widely adopted applications (e.g., \texttt{llama.cpp}~\cite{llamacpp}, Ollama~\cite{ollama}, SGLang~\cite{sglang}, ComfyUI~\cite{comfyui}) and popular models (e.g., Qwen3~\cite{qwen3}, Gemma3~\cite{gemma3}, Z-Image~\cite{zimage}, Qwen-Image~\cite{qwenimage}). Compared to application-managed solutions such as Ollama, as well as UVM and the UVM-based GPU multiplexing system nvshare~\cite{nvshare}, \sys reduces context-switch overhead by 29.1-82.3\% across configurations and cuts CPU pinned memory usage by up to 66.8\%. \sys delivers 1.3-1.6$\times$ speedups on diverse workloads that consists of using multiapplication workloads (e.g., using an LLM to generate prompts for image generation). \sys also improves interactive code-completion latency by 3.1-3.8$\times$, while maintaining competitive throughput on background batch-processing workloads.

Our paper makes the following contributions:
\begin{itemize}
    \item A new system architecture that jointly manages GPU memory usage and kernel dispatch, eliminating UVM-style thrashing without requiring any driver or application changes. The new architecture efficiently uses CPU pinned memory and bi-directional bandwidth between CPU and GPU.
    \item A lightweight, MLFQ-inspired scheduler for GPU workloads that automatically identifies and prioritizes interactive applications to reduce latency while preserving overall GPU throughput.
    \item An implementation that supports unmodified, state-of-the-art GPU applications (e.g., \texttt{llama.cpp}, SGLang, and ComfyUI) and delivers substantial improvements across diverse consumer GPU workloads.
\end{itemize}

\section{Motivation}
Here we motivate why the consumer GPU multiplexing problem is a fundamentally different problem from datacenter GPU multiplexing.

\subsection{Local ML Workloads on Consumer GPUs}

Recent advances in model quantization and optimized inference runtimes have enabled users to run large machine learning (ML) models directly on their personal machines. Applications such as local LLM inference (e.g., \texttt{llama.cpp}, SGLang, vLLM), text-to-image generation, and interactive image-editing pipelines (e.g., ComfyUI, diffusion-based tools) have become increasingly common. Users are motivated by privacy, offline capability, more fine-tuning opportunities and reduced hardware costs.

Unlike traditional datacenter GPU workloads, a single user may concurrently run or frequently switch between multiple large models: querying an LLM, generating an image, editing a photo, running background AI tasks, among other applications, all on a shared consumer-grade GPU. For interactive applications, the batch sizes almost always one.
At the same time, modern consumer GPUs (e.g., RTX 4090/5090) offer substantial compute but limited GPU memory relative to model size. Users often run the largest models that can fit in memory to maximize quality. However, this means each individual application may nearly saturate GPU memory.

\subsection{Limitations of Existing Multiplexing Approaches on Consumer GPUs}

\paragraph{Existing Approach \#1: Shut Down an Application and Run the Next Application.}
A straightforward solution is to run only one application at a time. For example, the community tool \texttt{llama-swap}~\cite{llama-swap} enables switching by terminating the current application and launching the next. However, unloading and reloading large models introduces multi-second delays, making even moderate task switching impractical for interactive consumer workflows (e.g., prompting an LLM to generate image descriptions that will be used immediately in a diffusion-based text-to-image application).

\paragraph{Existing Approach \#2: Application-Specific Approaches.}
Applications such as Ollama~\cite{ollama} provide model switching by unloading the current model and loading a new one. However, this mechanism is tightly coupled to a single application’s internal execution and cannot coordinate with other GPU applications. 

Another line of work makes an inference system’s GPU memory usage elastic. Prism~\cite{prism}, for example, allows an LLM framework to dynamically shrink its KV cache, freeing GPU memory for other applications. While effective for LLMs, this approach is highly domain-specific: diffusion models, image-editing pipelines, and many other consumer workloads do not use KV caches and therefore cannot benefit from this mechanism. Moreover, these systems require explicit software-level integration into each application or inference framework. As a result, they cannot provide transparent, system-wide multiplexing across the diverse set of applications on consumer GPUs.

\begin{figure}[t]
\centering
\includegraphics[width=0.95\linewidth]{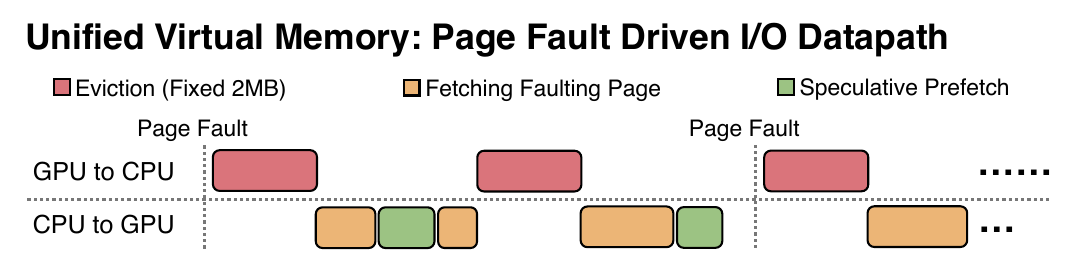}
\vspace{-3mm}
\caption{UVM's page fault driven migration. UVM can only utilize half link bandwidth for full-duplex PCIe channel.}
\label{fig:iopath}
\vspace{-5mm}
\end{figure}

\paragraph{Existing Approach \#3: NVIDIA Unified Virtual Memory (UVM).}

NVIDIA Unified Virtual Memory (UVM) is a fully transparent mechanism for handling GPU memory oversubscription. Its basic idea mirrors demand paging: when a GPU kernel accesses a page that is not resident in GPU memory, the hardware raises a page fault interrupt to the driver. The driver’s handler then selects pages for eviction, migrates data between GPU and CPU memory, and resumes kernel execution. UVM works well for: (1) a single application whose total memory footprint exceeds GPU capacity but whose working set fits in GPU memory, and (2) multiple applications whose combined working sets fit within GPU memory.

While UVM’s fully transparent abstraction is appealing, it has four fundamental limitations in the consumer setting.
First, UVM lacks coordinated control over compute and memory management, making it highly susceptible to thrashing. Consider two applications, A and B, running concurrently on a 32 GB GPU and generating one token each in a round-robin fashion. Suppose each model requires 24 GB of parameters; then each forward pass triggers the migration of at least 16 GB of data back to/from GPU memory. Even with state-of-the-art PCIe 5.0$\times$16 bandwidth (64 GB/s), this migration takes roughly 250ms. By comparison, a forward pass without migration typically takes only 20-75 ms. Thus, data movement alone induces a 4-12$\times$ slowdown.

Second, UVM utilizes only one direction of the PCIe bandwidth at a time, even though the PCIe channel is full-duplex. \autoref{fig:iopath} shows the IO path between CPU and GPU during page fault handling for UVM. During a page fault, UVM first evicts pages from GPU to CPU memory, and only after eviction completes does it fetch the demanded page and possibly several prefetched pages from CPU to GPU memory. A single fault may trigger multiple evictions followed by multiple fetches, but these transfers never overlap: eviction uses only GPU to CPU bandwidth, while fetching and prefetching use only CPU to GPU bandwidth. As a result, half of the available PCIe bandwidth remains idle.

Third, due to limited visibility into application behavior, UVM relies on an Least Recent Used (LRU)-based eviction policy to select pages during GPU page faults~\cite{demystifyinguvm}. However, UVM only updates its LRU metadata when page faults occur, which provides a highly incomplete view of an application’s true memory access pattern. As a result, pages that will be accessed imminently are often mistaken for cold pages, which leads to unnecessary page eviction and thrashing.

Fourth, host memory becomes a major bottleneck when UVM is used on consumer GPUs. To maximize PCIe transfer performance, UVM allocates CPU pinned memory inside the kernel for DMA operations. When UVM serves as the backing store for multiple GPU applications, every page resident on the GPU must have a corresponding pinned page on the CPU. This creates substantial memory overhead: even data that will never be evicted from the GPU still consumes unswappable, uncompressible CPU pinned memory. On consumer machines equipped with modest RAM, this CPU pinned memory footprint competes directly with other applications and can significantly degrade overall system usability.

Several GPU multiplexing systems build on top of UVM to mitigate some of its thrashing issues (e.g., TGS~\cite{TGS} and NVShare~\cite{nvshare}). For example, TGS assigns priorities to applications so that low-priority applications run only when high-priority ones are idle. 
However, because these systems rely on UVM as their underlying GPU memory virtualization mechanism, they inherit UVM’s fundamental limitations: page-fault-driven, single-directional PCIe transfers and large pinned-memory consumption.

\newcommand*\circled[1]{\tikz[baseline=(char.base)]{
  \node[shape=circle,draw,inner sep=1pt,fill=white,thick] (char)
  {\footnotesize\textbf #1};}}

\section{Overview}

Our design is guided by three key goals. First, we aim to provide a UVM-like abstraction that supports unmodified CUDA applications, while enabling coordinated control over compute and memory virtualization. Second, we seek to redesign virtual memory management to exploit a hierarchical memory system spanning GPU memory, CPU pinned memory, CPU pageable memory, and disk, while ensuring that CPU-GPU data transfers fully utilize the full-duplex PCIe bandwidth. Third, similar to standard OS kernel scheduler, we aim to optimize latency for interactive applications without sacrificing throughput for background batch-processing workloads.

\begin{figure}[t]
\centering
\includegraphics[width=0.99\linewidth]{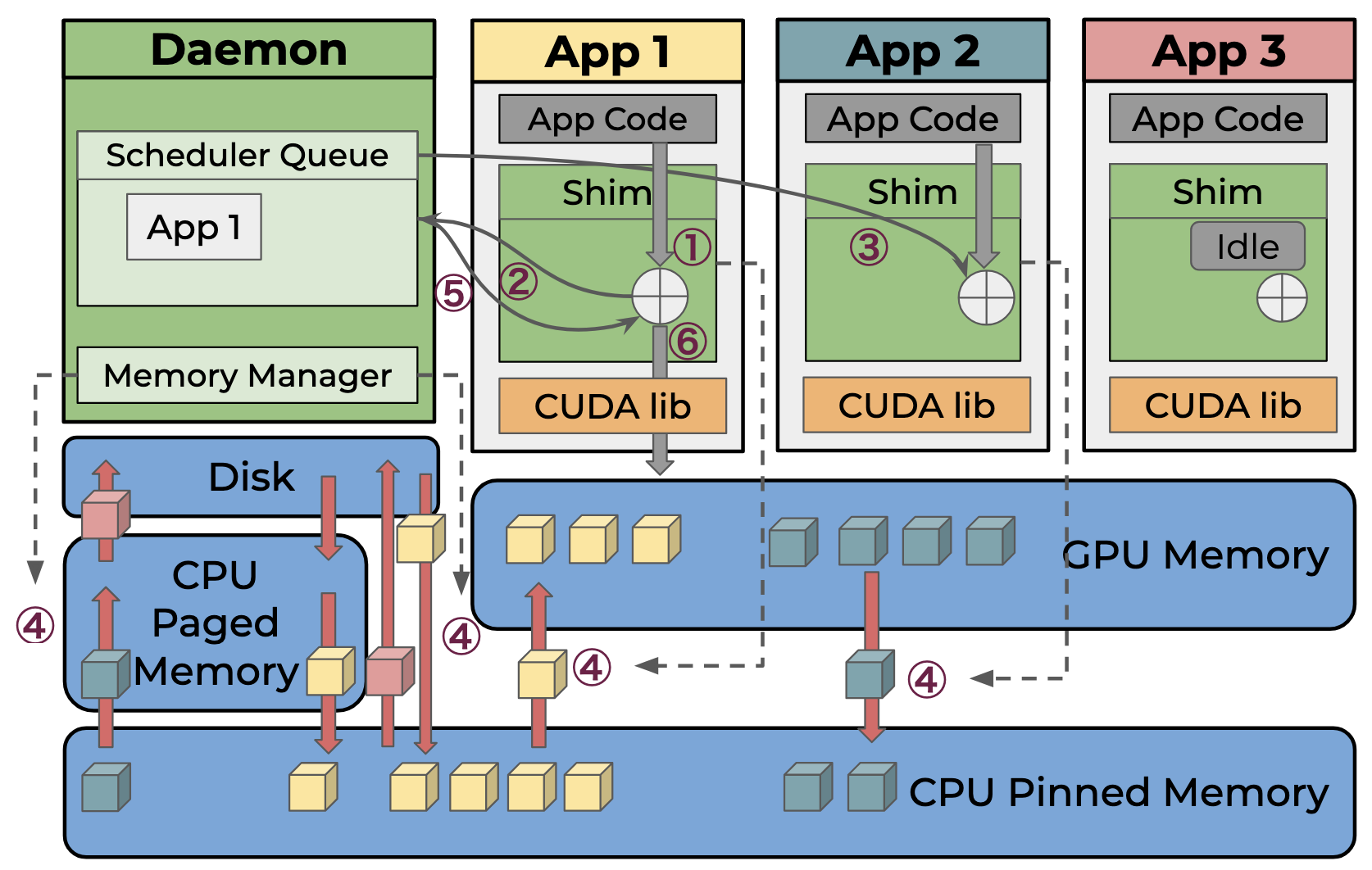}
\vspace{-3mm}
\caption{\sys's architecture overview. Green components belong to \sys. Cubes with different colors stand for memory blocks from different applications.} 
\label{fig:arch}
\vspace{-3mm}
\end{figure}

\autoref{fig:arch} presents the architecture of \sys. The system consists of two components: a lightweight application library, \shim, and a centralized system service, \daemon.
To provide transparency, \shim interposes on an application's CUDA Runtime API calls. \daemon performs global coordination and enforces multiplexing policies, but all kernels continue to execute using the application’s own CUDA context to minimize runtime overheads on kernel launches.
When App 1 launches a new kernel trying to execute GPU jobs, it \circled{1} checks the execution flag in the \shim; \circled{2} if App 1 cannot execute, \shim will then use IPC to enqueue the schedule request in the scheduler queue of \daemon. When the scheduler in \daemon decides App 1 is allowed to run, \circled{3} \daemon uses interprocess communication (IPC) to pause the execution of App 2, waiting for its completion of outstanding CUDA kernels. Then \circled{4} \daemon moves all the memory blocks belonging to App 1 into shared pinned memory, and each \shim migrates its own memory blocks into or out of GPU memory. After the memory blocks are ready for App 1, \circled{5} \daemon notifies \shim in App 1 to enable the execution flag; App 1 then \circled{6} resumes its execution. When all the data of App 1 is ready, App 1 only executes step \circled{1} and \circled{6}, bypassing IPC with \daemon. 

\daemon maintains a chunk-level virtual memory system. Instead of relying on page-fault-driven migration, \sys manages memory at the granularity of CUDA-allocated chunk (i.e., the units returned by cudaMalloc). Each chunk may reside in one of four locations: (1) GPU memory, (2) CPU pinned memory, (3) CPU paged memory, or (4) disk. Unlike UVM, chunks placed in GPU memory do not require CPU pinned memory. To avoid fragmentation in CPU pinned memory or paged memory, we split a chunk into 2MB blocks, where a block is the unit of data migration between the locations.

The scheduler in \daemon jointly manages (1) which application can launch kernels and (2) the locations of memory chunks in \sys. Its design follows two key principles. First, \sys coordinates the control over both compute and memory allocation.
When the scheduler selects an application to run, it proactively migrates that application's chunks into GPU memory while simultaneously evicting other applications’ chunks to CPU memory or disk. By coupling scheduling with memory placement, \sys ensures that when an application begins executing, it has both GPU compute and the necessary GPU-resident data available, avoiding demand-paging stalls and eliminating thrashing. Second, \sys prioritizes interactive workloads automatically. \sys determines whether an application is interactive by tracking its kernel launch behavior. Borrowing ideas from MLFQ, \sys adjusts each application's priority and time window dynamically. If an application does not issue a kernel during its assigned time window, \sys classifies it as interactive, halves its window size, and increases its priority. Conversely, if the application fully uses its assigned time window, \sys doubles the window size and lowers its priority. At each decision point, the scheduler selects the highest-priority application and grants it GPU access for its current window. This approach allows \sys to automatically prioritize latency-sensitive applications without any user annotations or application modifications, while preserving throughput for background batch processing applications.

\section{Supporting Unmodified CUDA Applications}
\sys is transparent to CUDA applications, similar to UVM. In UVM, a user replaces all the CUDA memory allocation from  \texttt{cudaMalloc} to \texttt{cudaMallocManaged}, and it is easy to make it fully transparent by static rewriting or \texttt{LD\_PRELOAD}.

\sys uses \texttt{LD\_PRELOAD} to hook the CUDA library and interpose several functions. Besides memory allocation and free, \sys also interposes CUDA kernel/graph launches in order to control the compute behavior. Further, \sys intercepts any APIs that implicitly allocate memory (e.g., \texttt{cudaStreamCreate}) and reports memory usage (e.g., \texttt{cudaMemGetInfo}). Applications sometimes require correct memory usage reporting to function correctly. For example, SGLang reads GPU memory in order to decide how much KV cache memory to be allocated. \shim provides such interceptions and also communication to \daemon via UNIX domain socket.

To make sure applications can run correctly without being aware of the existence of \sys, we need to keep an identical running environment before pausing and after resuming, and no invalid GPU operations should be introduced by \sys.

\paragraph{Maintaining consistent virtual memory addresses.} Classic approach like \texttt{cudaMalloc} and \texttt{cudaFree} cannot guarantee the virtual memory addresses are consistent after another call. Applications which use the previous addresses will access memory regions that are no longer valid. \sys leverages CUDA VMM API\cite{nvidia_cuda_va} to reserve GPU virtual addresses, and map these addresses to different physical allocations before and after chunk migrations.

\paragraph{Valid memory access during migration.} GPU kernel execution is mostly asynchronous. When migration starts, there can be outstanding kernels still executing, and invalid memory accesses can happen when the accessed pages have been migrated. Before the current running application is being migrated, \sys blocks any new kernels from launching and performs a CUDA synchronization to complete all the outstanding kernels. This prevents an application from using the migrated chunks.

\paragraph{CUDA graph compatibility.} CUDA graph is widely used to reduce kernel launch overhead. When applications are constructing CUDA graph, unexpected API calls can disrupt the graph capture and lead to unexpected behavior. \sys ensures \shim does not call any CUDA APIs until the graph capture finishes by interposing on the start and the end of capture events.

\paragraph{GPU Memory information.} Some applications assume they have exclusive access to the entire GPU, and use device-wide statistics to make decisions for tensor placements or buffer allocations. \sys reports GPU memory usage by the sum of allocations from the process itself, preventing applications from unnecessarily spilling data to CPU.

\section{Virtual Memory Management}
We first describe the granularity of memory management in \sys, and then describe how to migrate memory chunks within hierarchical memory.

\subsection{Granularity of Memory Management}
\sys intercepts all GPU memory allocations in applications, and it is thus naturally aware of all  memory allocations.
\sys manages memory at the granularity of each allocation and migrates these units as a whole, which we refer to as \textit{chunks}. Each chunk is placed at a specific tier of the memory hierarchy. Some applications (e.g., \texttt{llama.cpp}) put all the model weights into one big allocation. To prevent excessively large chunks, \sys caps each chunk’s size at 128 MB. Larger allocations are subdivided accordingly, allowing portions of the same allocation to reside in different memory hierarchies. For example, a 256 MB allocation may be split so that roughly half resides in CPU pinned memory while the remainder is placed in CPU paged memory.

Allocations naturally vary in size because tensors (e.g., weights, activations) have diverse shapes. If \sys managed memory solely at the chunk level, this variability would lead to severe fragmentation. To address this, \sys introduces blocks for fine-grained control. Each block has a fixed size of 2 MB, which is the smallest physical allocation unit supported by the VMM API. A chunk is composed of one or more blocks (up to 64). Using uniformly sized blocks simplifies placement, migration, and bookkeeping, enabling \sys to manage memory efficiently while avoiding fragmentation.

\begin{figure}[t]
\centering
\includegraphics[width=0.99\linewidth]{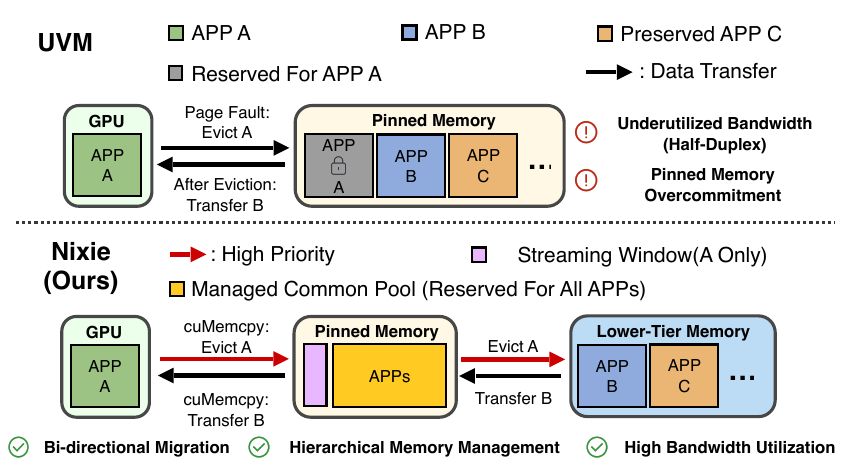}
\vspace{-3mm}
\caption{\sys's memory hierarchy and chunk migration.} 
\label{fig:hierarchy}
\vspace{-3mm}
\end{figure}

\subsection{Hierarchical Memory}

One typical implementation of this hierarchical memory is to pin all the memory of tier $X$ at tier $X+1$, and view tier $X$ as a caching layer for tier $X+1$. This works great if tier $X+1$ always has significantly more capacity than tier $X$. This is a valid assumption in CPU cache hierarchy, and is the assumption of UVM in terms of viewing GPU memory as a cache for CPU pinned memory.

However, this assumption breaks down on consumer machines, where CPU pinned memory is limited and often comparable in size to GPU memory. Consider three applications, each requiring 24 GB of GPU memory: UVM would need 72 GB of CPU pinned memory to serve as the backing store. While pinned memory offers high transfer throughput, it cannot be swapped out or transparently compressed by the operating system. This is manageable on dedicated inference servers running a single workload, but on consumer hardware, where many applications coexist, excessive pinned-memory usage can severely degrade overall system usability.

Instead of making tier $X$ as a caching layer of tier $X+1$, we ensure that each chunk has \textit{exactly one copy} within the entire memory hierarchy and is migrated between tiers as needed. \autoref{fig:hierarchy} shows the \sys's memory hierarchy. \daemon is the only component that issues data migration and has all the information needed to decide chunks' locations in the memory hierarchy. \daemon can proactively perform background migrations without involving or interrupting applications. This allows \sys to use much less pinned memory compared to UVM.

\begin{figure}[t]
\centering
\includegraphics[width=0.99\linewidth]{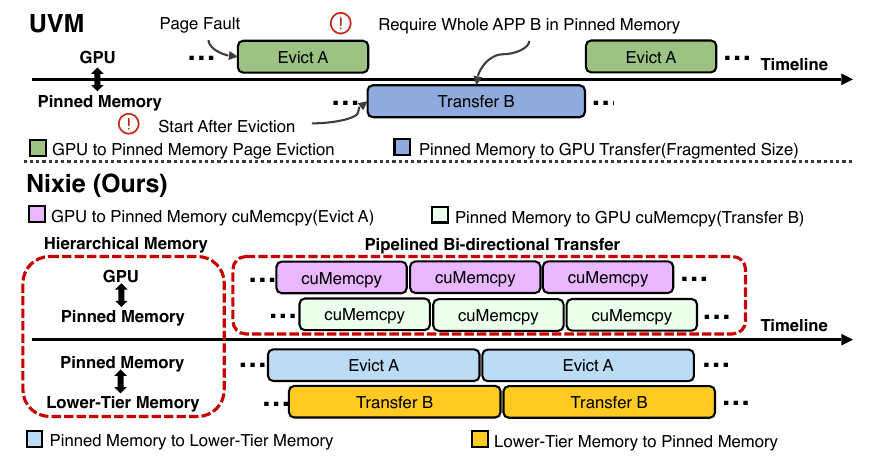}
\caption{\sys's communication bandwidth usage (1) between GPU memory and CPU pinned memory, and (2) between CPU pinned memory to lower-tier memory (e.g., CPU paged memory, disk).} 
\label{fig:migration}
\vspace{-3mm}
\end{figure}

\subsection{Speeding Up Context Switches}
Migration is needed when the GPU memory is not enough to fit the next application to run, and long migration time can severely affect user experience. To make the duration of migration as short as possible, we need to fully utilize the link bandwidth. We demonstrate the migration process in \autoref{fig:migration}, where the upcoming application B needs to migrate data into the GPU and the current application A needs to be evicted.

A straightforward solution is to create multiple bi-directional transfer tasks across the nearby tiers individually, and use back-pressure to make sure the space is available for transfer. However, in the multi-tier hierarchical scenario, the back-pressure chain is long. To satisfy the transfer request from upstream, the back pressure will be passed down multiple times, and the transfer from the upstream can only happen after corresponding the downstream finish. This can incur a long latency and reduce the utilization of the transfer links. Another issue is contention, especially when the capacity of the current tier is smaller than the size of the application. When transfer is not coordinated, it is possible that the data from B fully occupies this tier. In this case, we will either have to move the data back to a lower tier, creating unnecessary traffic, or a more severe deadlock when we expect the data from B only go to upper layers.

\begin{figure}[t]
\centering
\includegraphics[width=0.99\linewidth]{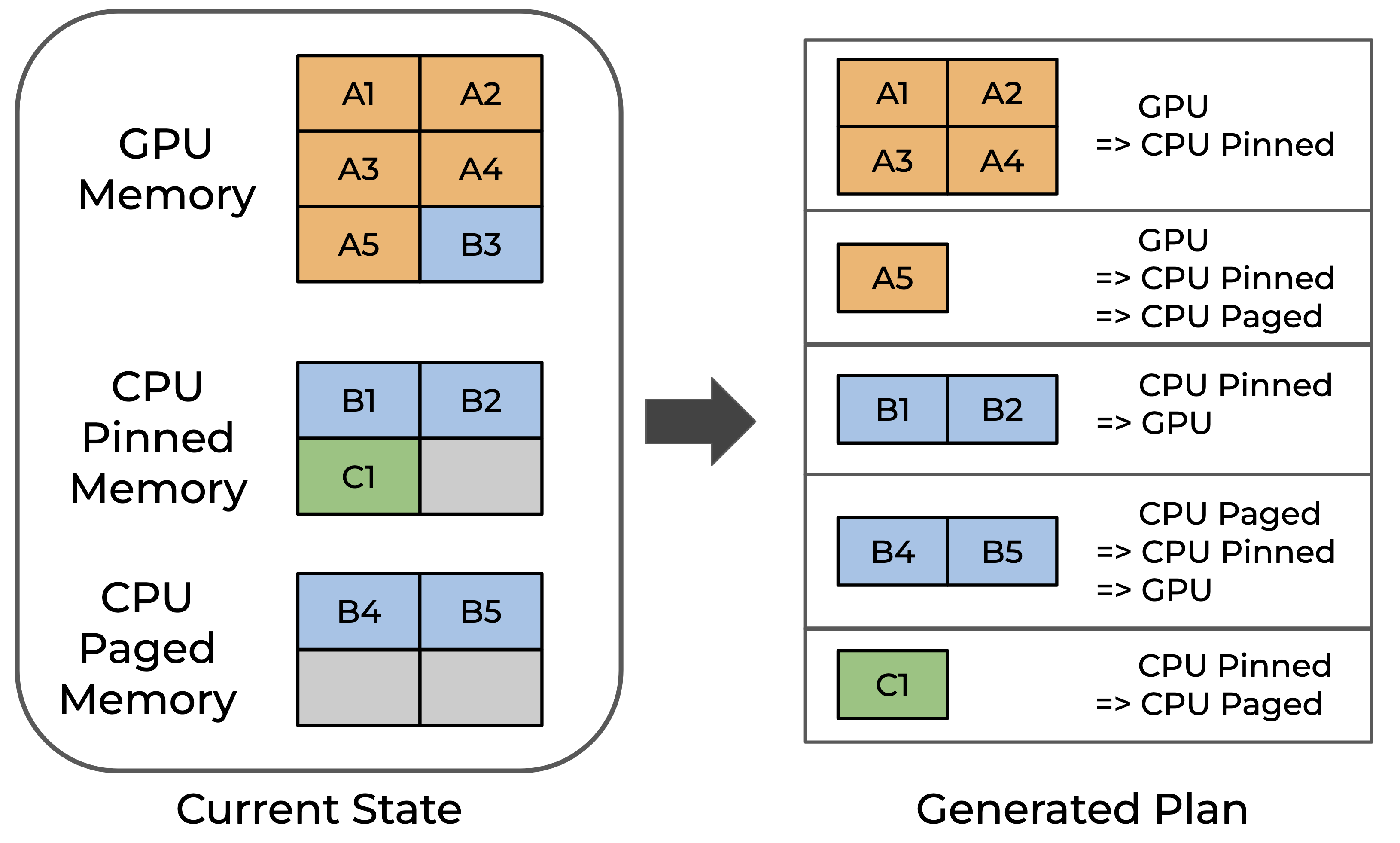}
\vspace{-3mm}
\caption{Migration plan given existing chunks' locations.} 
\label{fig:plan}
\vspace{-3mm}
\end{figure}

To resolve these problems, we implement two solutions. 
First, instead of directly starting the transfer and purely resolving contention locally, we make a global plan by inspecting the current usage of all applications. The planner decides what data chunk needs to be migrated, and its source and destination as shown in ~\autoref{fig:plan}. \daemon collects all the metadata of chunks needed to be migrated to GPU, and calculates what data chunks on GPU should be evicted. After the eviction decision is made, the planner goes through tiers to determine how much evicted data can be kept in that tier without being further transferred to lower tiers. The planner prioritizes transferring that crosses multiple tiers.

Second, we add a migration orchestrator in \daemon to manage the migration order and priority given the limited space and the different chunk source and destination.  We also reserve a small streaming window dedicated to the evicting application A. By reserving it, we make sure the downstream of the back-pressure chain not be blocked. Also, when there is no back pressure, the data from A can safely stay here without being further moved and without harming latency. 

For the migration, \daemon moves the necessary data of the incoming application to the CPU pinned memory. Multi-threading transfer is used for saturating the DRAM bandwidth. When the incoming data is ready on the CPU pinned memory, the ID of chunks and their data blocks are sent to \shim via IPC. \shim leverages asynchronous memcpy to transfer data from 2MB blocks to and out of GPU.

\section{Scheduling}
Similar to traditional CPU scheduling, our goals are: (1) interactive applications should receive higher priority than batch-processing applications, and (2) users should not be required to manually specify application priorities. On CPUs, multi-level feedback queues (MLFQ) are the canonical mechanism to achieve both goals.

GPUs, however, introduce new challenges. First, there is no direct signal indicating whether an application is interactive or batch-processing. On CPUs, interactivity can be inferred through events such as voluntarily yielding the CPU (e.g., via \texttt{yield()}) or issuing blocking system calls for I/O, but such signals do not exist in GPU workloads. Second, GPU context switching incurs significantly higher overhead due to the need to migrate large memory chunks across memory hierarchies.

In the following, we describe how \sys detects applications not using GPUs and then infers application priority. Finally, we discuss how to use prefetching to augment an MLFQ-like scheduling system tailored for GPU workloads.

\subsection{Idleness Detection}

In traditional CPU scheduling, a scheduler is typically invoked when (1) a time window expires or (2) a thread voluntarily relinquishes the CPU, for example by issuing \texttt{yield()} or invoking a blocking system call (e.g., I/O). These signals allow the CPU scheduler to switch to another thread of execution. However, neither signal exists in GPU applications. First, \sys needs to support applications with long-running services (e.g., SGLang, Ollama), so there is no notion of ``task completion.'' Second, GPU applications do not expose equivalents of yield or blocking system calls to indicate periods of inactivity. As a result, letting applications run until the time window expires, regardless of whether an application is doing useful work, leads to poor GPU utilization and high latency for interactive workloads.

To address this, we design a mechanism to detect when a GPU application becomes idle. A straightforward approach is to observe GPU utilization using device APIs such as NVML. Unfortunately, this has two drawbacks. First, GPU-level APIs are coarse-grained and update slowly. In our experiments, NVML takes roughly 600\,ms to report that a workload has become idle. Second, On consumer devices, background tasks (e.g., rendering or video decoding) introduce noise that makes these signals unreliable.

We use a timeout-based mechanism to detect application idleness. CUDA APIs fall into two categories: non-blocking operations (e.g., kernel launches, asynchronous calls) and blocking operations (e.g., synchronization). \sys marks an application as idle only if the time elapsed since its most recent API return exceeds a fixed threshold. For blocking APIs, \shim intercepts the call both before and after execution to ensure the application is not mistakenly classified as idle while it is blocked inside the runtime. In our implementation, we use a 100 ms threshold. 

\subsection{Priority Inference}

Classic MLFQ decreases priorities over time and periodically resets all priorities to the highest level. This works well on CPUs, where time windows are short. In \sys, however, GPU time windows are much longer. Blindly resetting all priorities erases useful history and may unexpectedly preempt interactive workloads.

We introduce a soft priority recovery policy. The scheduler records three timestamps for each application $a$: (1) the time it became idle $i_a$, (2) the time of last priority update $p_a$, and (3) the time of its most recent scheduling request $q_a$. Ideally, when an application stays idle for a sufficiently long period, it indicates the usage pattern may have changed (e.g., a LLM generating a long response for one prompt now may generate a short response for another prompt), so its priority should be increased to re-estimate the runtime characteristics. However, if the application is already in the scheduler queue pending for further execution, promoting it immediately would preempt the currently running application and reduce the effective time window of every application in that priority level. 
To address this issue, we exclude pending time from consideration. Let $R$ denote the multiplying factor applied to the pending time. If $R \geq 1$, an application could potentially starve at low priority whenever high-priority jobs are continuously running. Therefore, we dynamically adjust $R$ based on the number of applications $N$ at the same priority level, ensuring that $R<1/N$. The detailed priority inference algorithm is shown in \autoref{alg:schedule}.

\begin{algorithm}[t]
\caption{\sys Priority Inference}
\begin{algorithmic}[1]
\Require Application $a$, pending time factor $R$, set of multi-level feedback queues $\{Q\}$ and corresponding time allotment $\{T\}$
    \State $p \gets$ current priority of $a$
    \State $t_a \gets$ accumulated execution time of $a$ in current priority
    \If{$t_a > T_p$} \Comment{Standard MLFQ Demotion}
        \State Move $a$ to lower priority queue
        \State $t_a \gets 0$
    \ElsIf{$a$ is idle}
        \State $i_a \gets$ time since last execution
        \State $p_a \gets$ time since last priority change
        \State $q_a \gets$ time since $a$ enqueued (0 if no new request)
        
        \If{$p_a > T_p$} \Comment{Prevent priority jitter}
            \State \emph{// Compensate idleness and prevent starvation}
            \If{$i_a - R \cdot q_a > T_{p-1} + t_a$}
                \State Move $a$ to higher priority queue
            \EndIf
        \EndIf
    \EndIf

\end{algorithmic}
\label{alg:schedule}
\end{algorithm}

\subsection{MLFQ with Prefetching} \label{sec:prefetch}

\sys uses K queues $(Q_1,Q_2,\ldots ,Q_k)$ where the priorities are in descent order from $Q_1$ to $Q_k$. The scheduler prioritizes applications with higher priority levels; when multiple applications share the same priority, they are scheduled in round-robin fashion, beginning with the application that has waited longest. Upon reaching a preemption threshold $S$, the scheduler preempts the current application if another application exists at the same priority level. We set the time allotment for priority demotion at $T=8\text{s}$ and the preemption threshold at $S=4\text{s}$
for the highest priority queue; each successive lower-priority queue doubles these parameters. The scheduler periodically checks for new scheduling decisions and updates priorities. Upon receiving an idleness notification from the currently running application, the scheduler performs the same operations.

In \sys's hierarchical memory design, moving data from a lower to a higher tier requires a long time. We leverage the information from scheduler to reduce the migration time: when there is already an application at the head of scheduler queue, it is most likely to run after the current application being scheduled out. \sys thus prefetches data belonging to the next application to reduce the migration time. 

\section{Evaluation}

We have prototyped \sys using \textasciitilde10,000 lines of Rust code. \shim has \textasciitilde1900 lines of code, and \daemon has \textasciitilde7400 lines of code, with \textasciitilde700 lines of shared code. Next, we microbenchmark \sys’s context-switching latency as well as its compute and memory overheads. We then evaluate a set of representative use cases to demonstrate \sys’s practical benefits.

\paragraph{Setup.} Our experiments are conducted on a desktop machine equipped with an AMD Ryzen 9 9950X CPU and 96 GB of dual-channel DDR5 memory at 3600MT/s. The system includes two 32GB RTX 5090 GPUs, each connected via PCIe 5.0x8. The desktop machine runs Debian 12 with CUDA version 12.9 and NVIDIA driver version 580.95.05. We run ML models using Ollama~\cite{ollama} (version 0.12.11), SGLang~\cite{sglang} (version 0.5.4.post1), \texttt{llama.cpp}~\cite{llamacpp} (version b7027) and ComfyUI~\cite{comfyui}(commit eaf68c9). We pick a variety of models from Qwen3\cite{qwen3}, Gemma3\cite{gemma3}, Z-Image\cite{zimage} and Qwen-Image\cite{qwenimage} with different sizes and quantizations. 
For example, Qwen3-MoE 30B-Q6 means it is the 30B variant of Qwen3-MoE with 6-bit quantization. 

\paragraph{Baselines.} We compare \sys with Ollama~\cite{ollama}, UVM~\cite{uvm}, TGS~\cite{TGS}, and nvshare~\cite{nvshare}. The UVM baseline is implemented by hooking only \texttt{cudaMalloc}, \texttt{cudaFree}, and \texttt{cudaMemGetInfo} three APIs. We include Ollama as both an application atop \sys and as a baseline: Ollama executes ML models and can switch between models natively. We only evaluate TGS using cases studies and exclude TGS from microbenchmarks because TGS only supports exactly two applications and requires users to explicitly designate one as high priority and the other as low priority, making it incompatible with our microbenchmark workloads.

\subsection{Microbenchmarks}

\begin{figure}[t]
\centering
\includegraphics[width=0.95\linewidth]{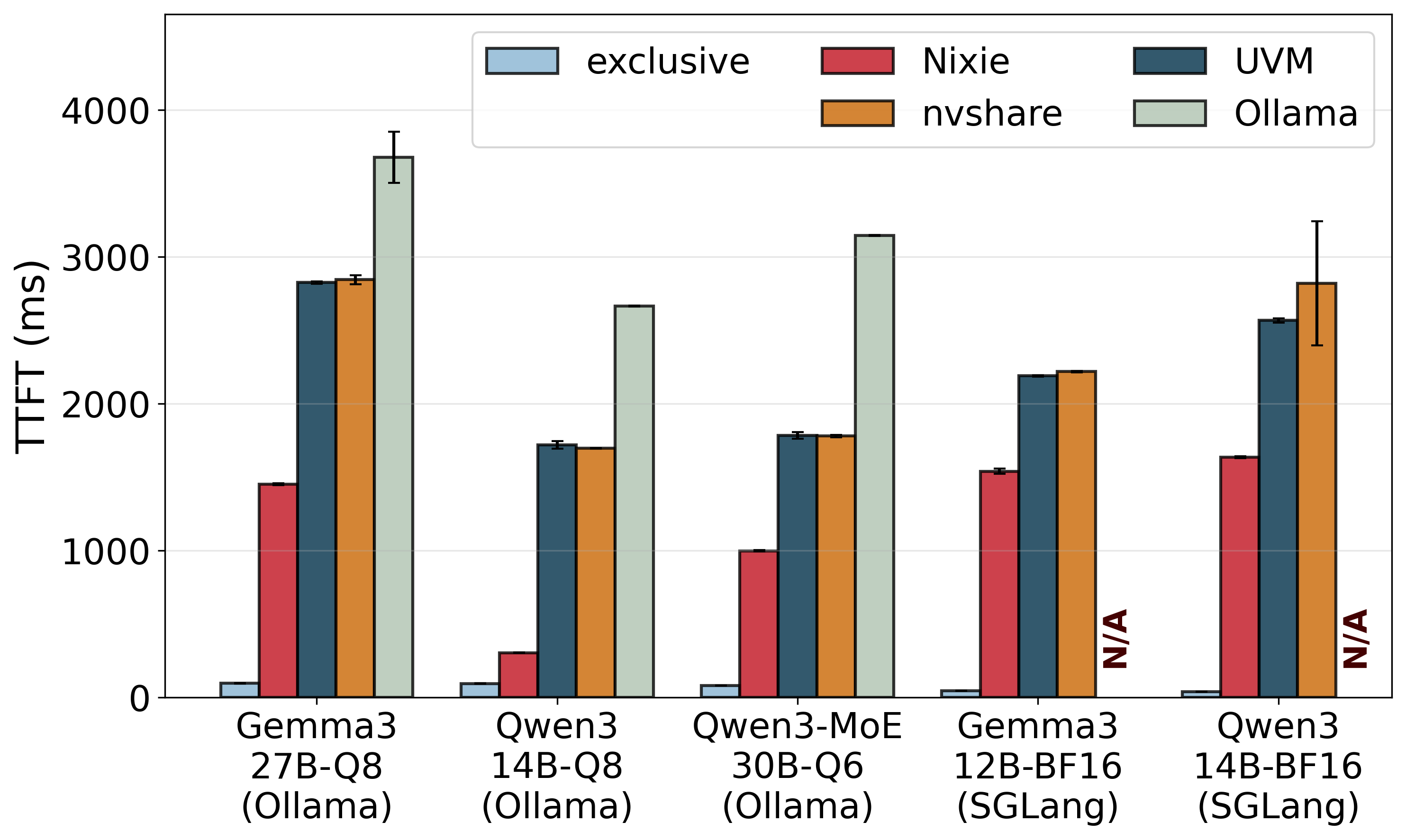}
\vspace{-3mm}
\caption{Context switch performance comparison. Error bars represent standard deviations.}
\label{fig:rotate_ttft}
\vspace{-3mm}
\end{figure}

\begin{figure}[t]
\centering
\includegraphics[width=0.95\linewidth]{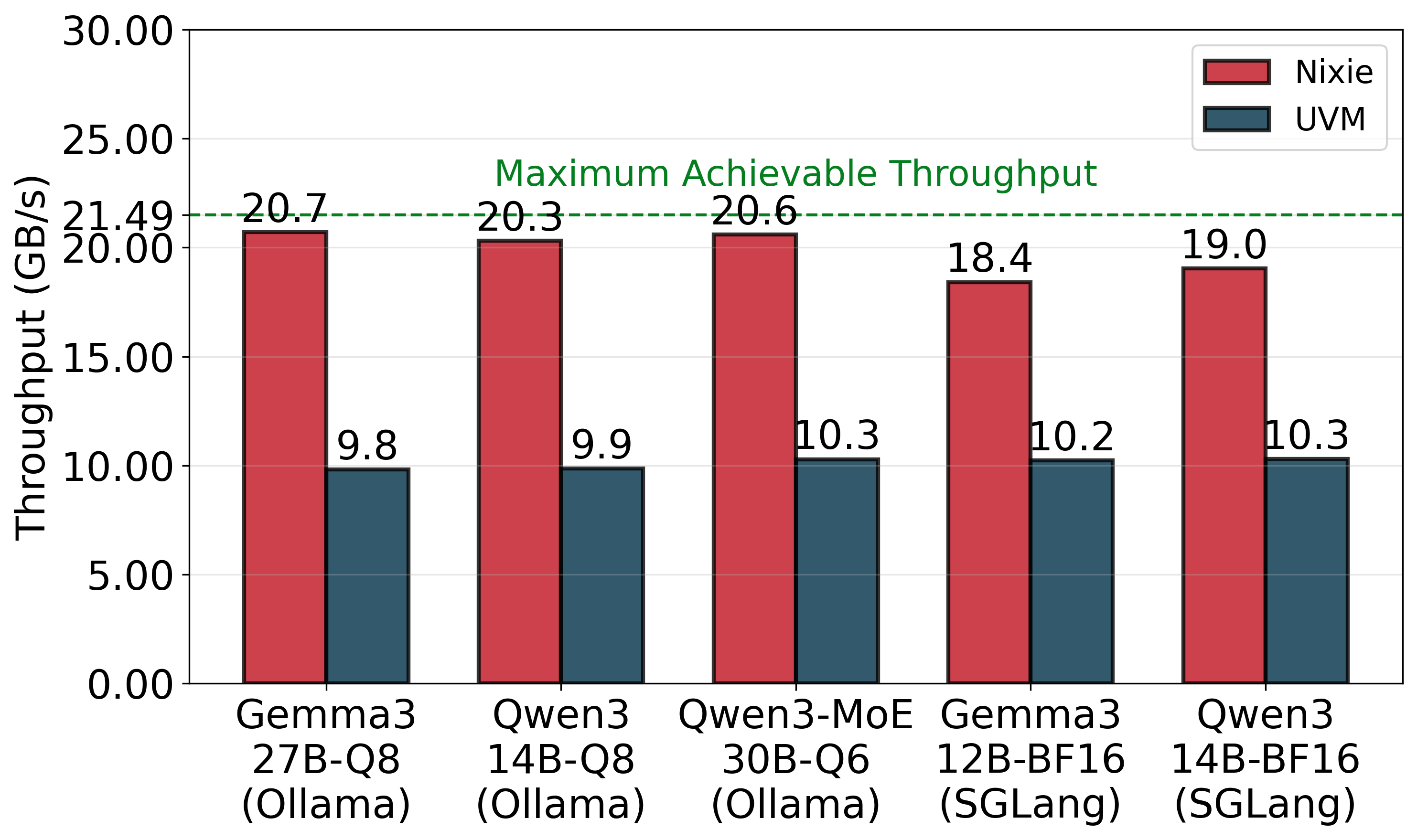}
\vspace{-3mm}
\caption{CPU from/to GPU memory copy throughput.}
\label{fig:tput}
\vspace{-5mm}
\end{figure}

\paragraph{Context Switch Performance.}
We evaluate context-switch performance by running an application executing a model and then switching to another identical instance of the same application and model on a single GPU. \autoref{fig:rotate_ttft} reports the Time-To-First-Token (TTFT) under five settings: \emph{exclusive}, \sys, UVM, nvshare, and Ollama. Here, \emph{exclusive} places the two applications on separate GPUs, and thus reflects TTFT assuming the context switch is effectively free.

Overall, \sys reduces TTFT by 44.0\%-82.3\% for Ollama cases and 29.7\%-36.3\% for SGLang cases compared to UVM and nvshare. Ollama uses a simple \texttt{mmap}-based strategy to unload and reload models, which results in the worst performance among all solutions tested. 

We highlight two interesting observations. First, for MoE models (e.g., Qwen3-MoE), the performance benefit of \sys for TTFT is smaller because the prefill only activates a subset of experts. Second, \sys provides larger benefits for Ollama than for SGLang. SGLang aggressively allocates KV-cache memory that remains unused; \sys proactively offloads all unused KV-cache pages to CPU memory, while UVM moves such pages in a delayed manner.

To understand why \sys shortens context switch time compared to UVM and nvshare, we measure the bi-directional data-transfer throughput between host and GPU memory. \autoref{fig:tput} shows the results. For reference, the dotted line marks the maximum achievable bi-directional throughput on our testbed, measured by NVIDIA’s \texttt{nvbandwidth}~\cite{nvbandwidth} tool. 

\sys nearly saturates the available bi-directional bandwidth and achieves around 2$\times$ the throughput of UVM. The remaining gap between \sys and the theoretical maximum primarily comes from memory allocation overheads. Ollama exhibits a simple, large size allocation pattern, whereas SGLang has a more complex and non-uniform size pattern. Consequently, \sys is able to reach higher bi-directional throughput when running Ollama than when running SGLang.

\begin{figure}
\centering
\includegraphics[width=0.95\linewidth]{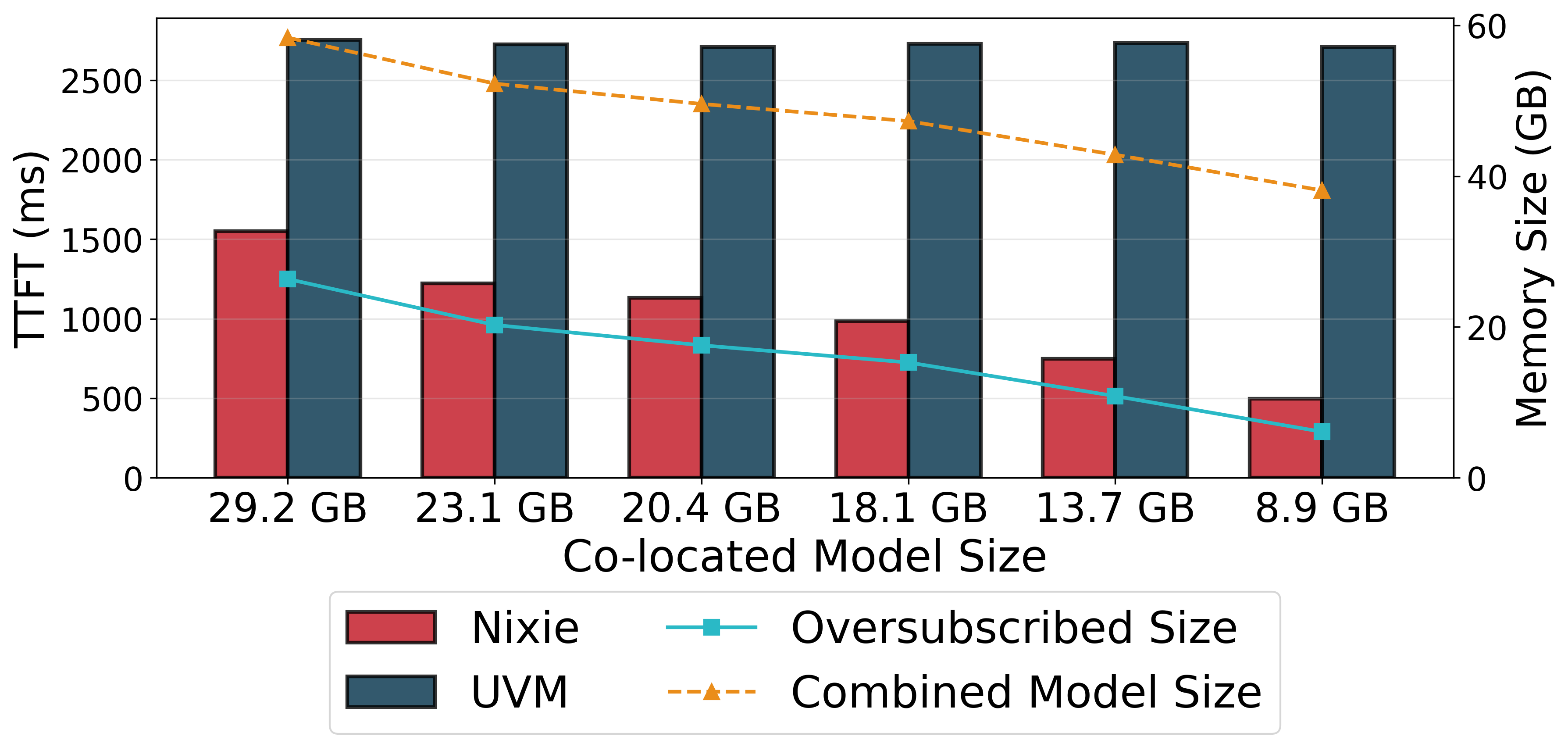}
\vspace{-3mm}
\caption{Performance comparison under different amounts of memory oversubscription. Oversubscribed size = Combined model size - GPU memory size.}
\label{fig:trend}
\end{figure}

In \autoref{fig:trend}, we analyze the effect on TTFT when co-locating with applications of varying sizes. We use \texttt{llama.cpp} as our primary application and Gemma3 27B-Q8 as the model for measuring TTFT. As total GPU memory oversubscription decreases, \sys achieves improved context switch performance, whereas UVM maintains constant context switch overhead due to its LRU policy, regardless of the extent of GPU memory oversubscription.

\begin{figure}[t]
\centering
\begin{subfigure}[b]{0.95\linewidth}
\includegraphics[width=0.95\linewidth]{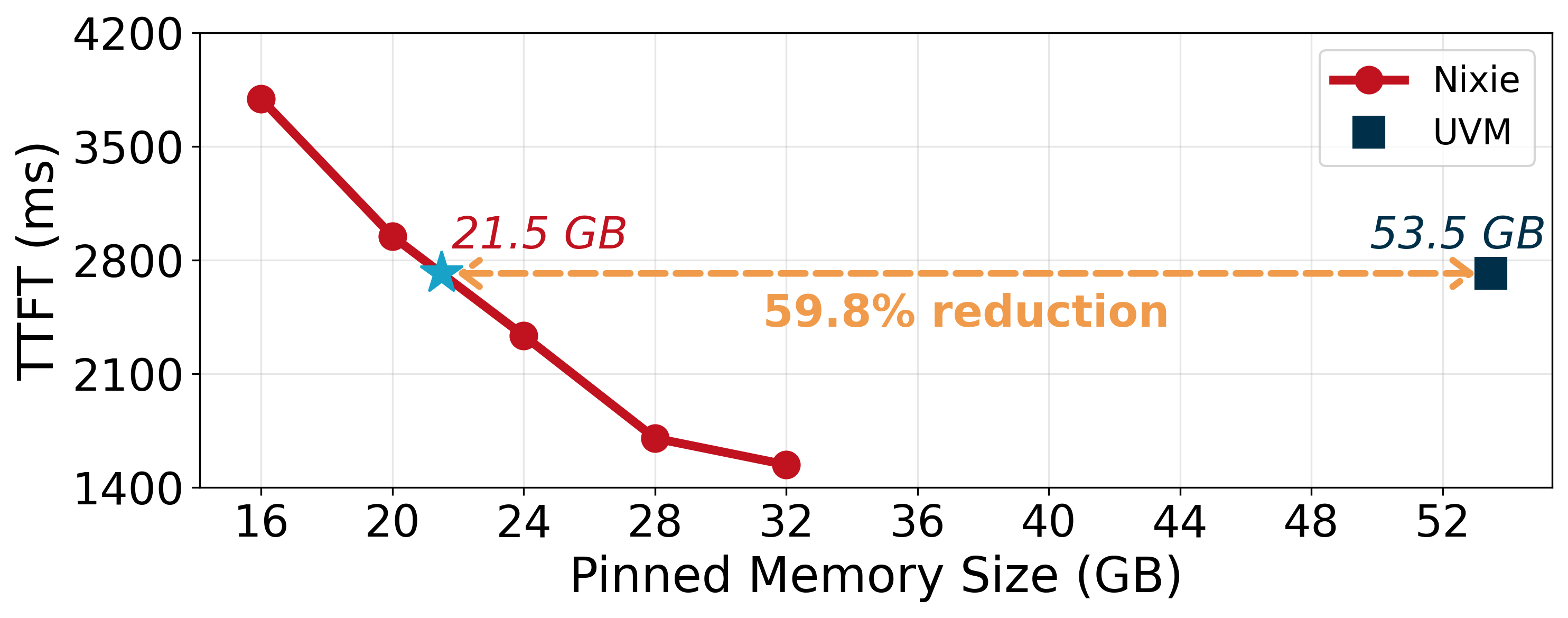}
\caption{Gemma3 27B-Q8}
\end{subfigure}
\begin{subfigure}[b]{0.95\linewidth}
\includegraphics[width=0.95\linewidth]{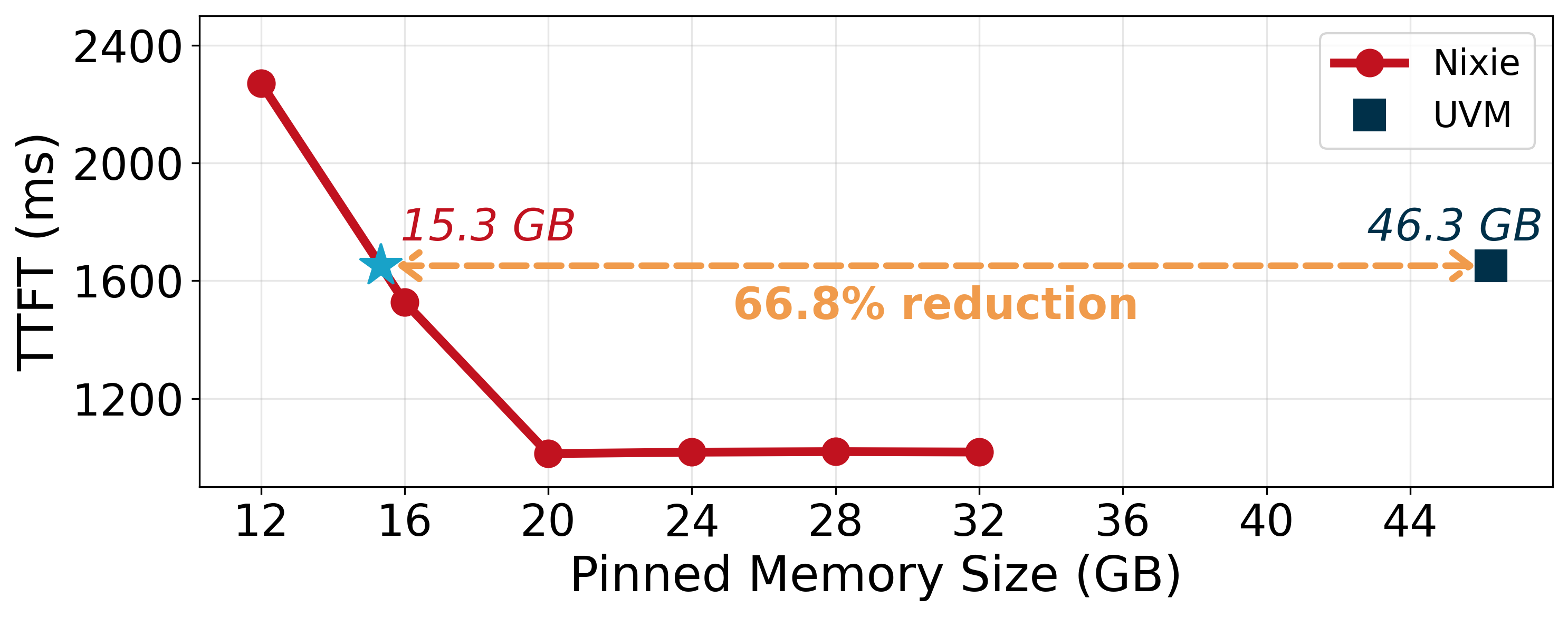}
\caption{Qwen3-MoE 30B-Q6}
\end{subfigure}
\vspace{-3mm}
\caption{Pinned Memory Versus TTFT}
\label{fig:space}
\vspace{-5mm}
\end{figure}

Sometimes, users may have insufficient CPU memory when they have to work with other applications. UVM leverages pinned memory for DMA to achieve best migration performance. We vary the total amount of CPU pinned memory \sys can use from 16GB to 32GB. In \autoref{fig:space}, we show that \sys can achieve the same performance of UVM with only 33.2\% to 40.2\% of CPU pinned memory used in UVM. 

\begin{figure}[t]
    \centering
    \begin{subfigure}[t]{0.48\linewidth}
        \centering
        \includegraphics[width=\textwidth]{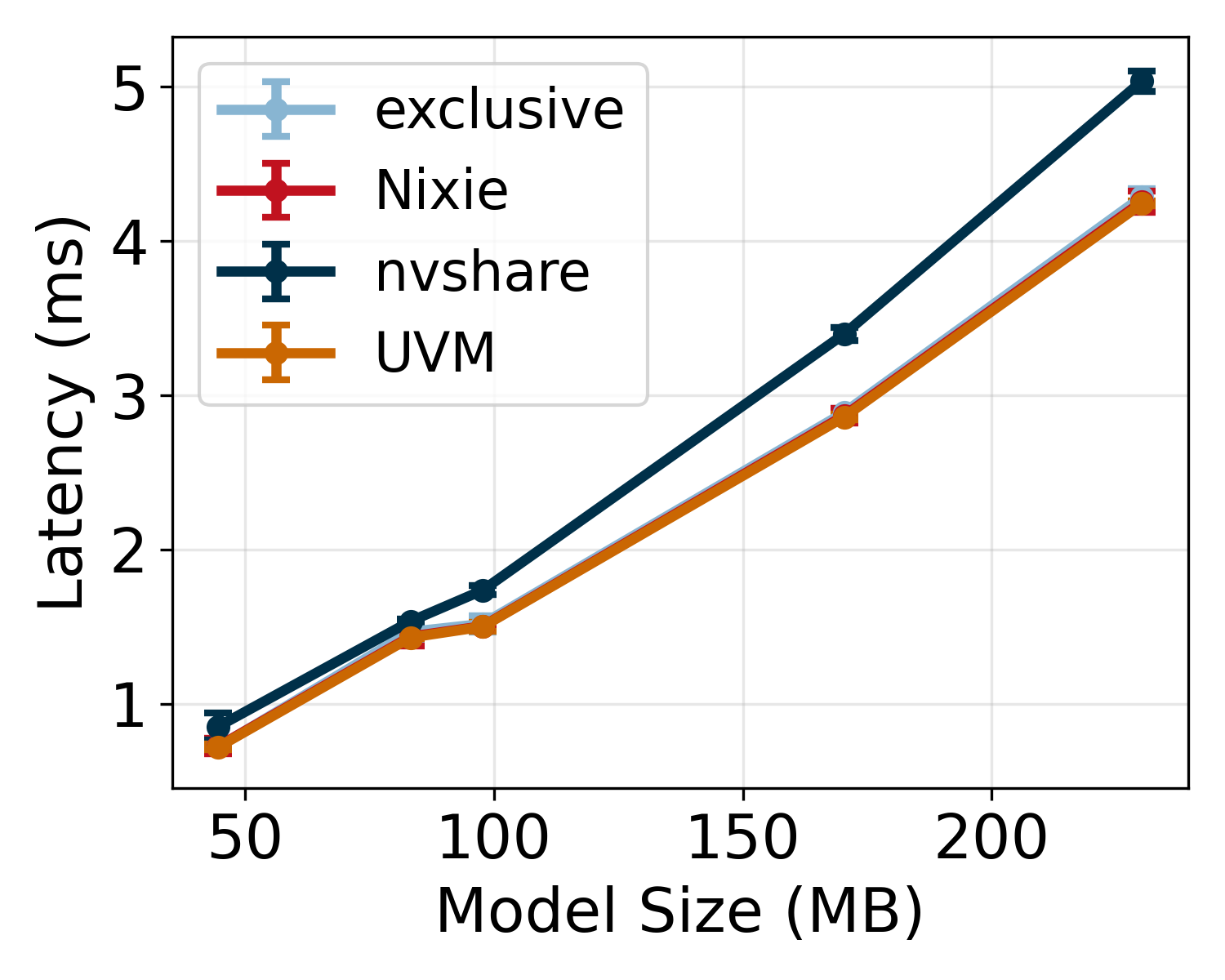}
        \caption{Inference latency}
        \label{fig:resnet_overhead}
    \end{subfigure}
    \hfill
    \begin{subfigure}[t]{0.48\linewidth}
        \centering
        \includegraphics[width=\textwidth]{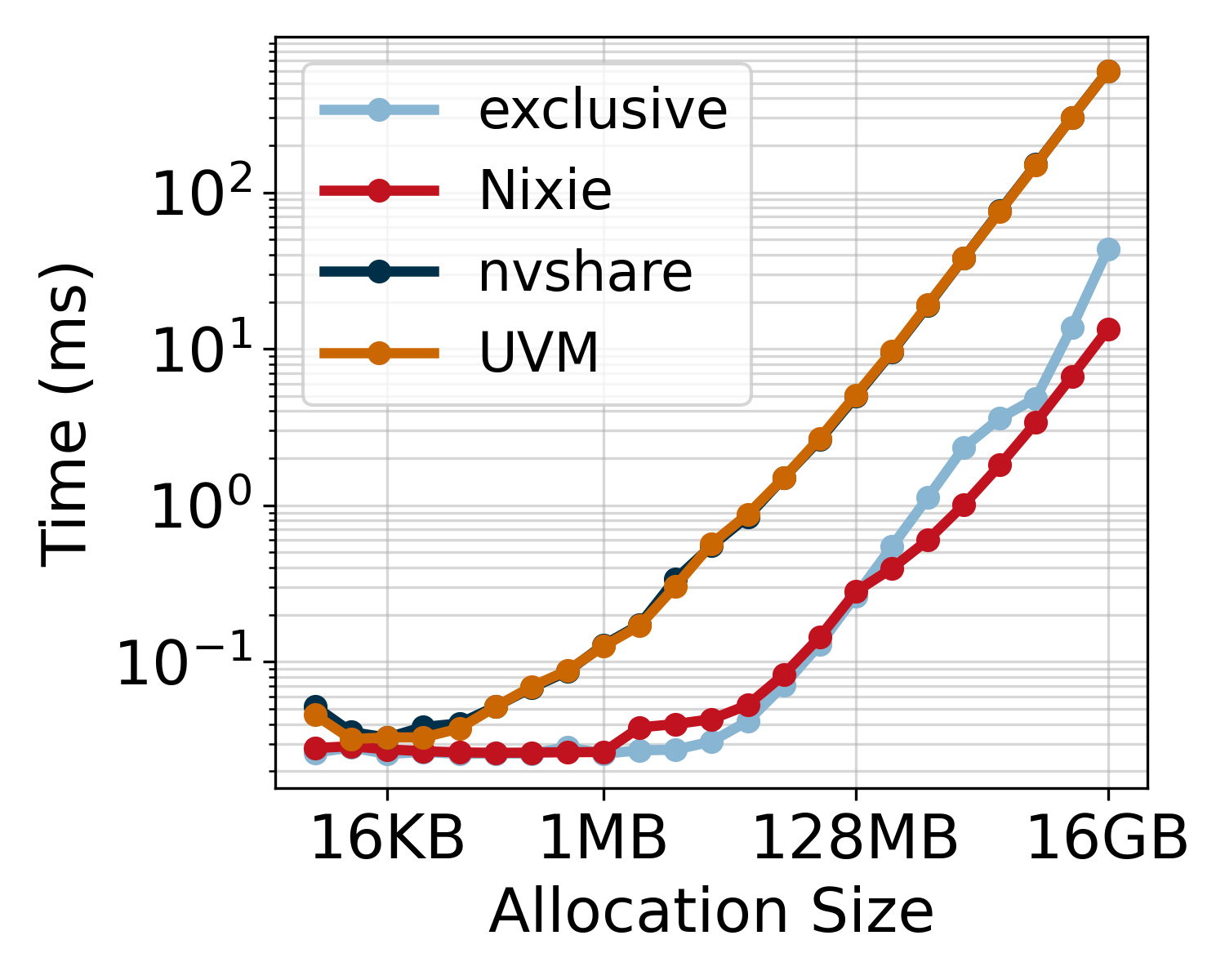}
        \caption{cudaMalloc latency}
        \label{fig:malloc_overhead}
    \end{subfigure}
    \caption{\sys's runtime overheads.}
    \label{fig:overhead}
    \vspace{-3mm}
\end{figure}

\paragraph{Runtime Overheads.}
During the time an application is scheduled to run on a GPU, \sys should have minimal performance overheads on kernel launch, even when kernel itself is small. This is because during the window the application is scheduled to run, the CUDA launch calls directly go through the corresponding CUDA userland driver and do not need to be handled via \daemon.

\autoref{fig:resnet_overhead} displays inference latencies of various small ResNet models ranging from ResNet-18 to ResNet-152 at a batch size of 1. \sys shows the identical performance with vanilla execution from \emph{exclusive} and simple UVM interception. The overhead from nvshare mainly comes from its CPU thread synchronization and GPU synchronization under certain conditions.

\paragraph{Memory Overheads.}
In \autoref{fig:malloc_overhead}, we measure the latency of \texttt{cudaMalloc()} with varying allocation sizes. We execute a subsequent \texttt{cudaMemset()} to eliminate the effects of lazy initialization. For allocations below 2MB, \sys directly uses \texttt{cudaMalloc()}, achieving performance comparable to the vanilla implementation. Between 2MB and 128MB, \sys exhibits slightly higher allocation times compared to the vanilla implementation. Because the CUDA runtime is closed-source, we can only hypothesize why \sys outperforms the vanilla implementation once the allocation size exceeds 128 MB. Our hypothesis is that \sys allocates memory using multiple contiguous 128 MB physical regions, whereas the vanilla implementation may attempt to allocate a single, maximally contiguous region. UVM and UVM-based nvshare shows worse performance because UVM adopts a first-touch allocation policy, which defers physical memory allocation until the initial access, introducing more overhead.

\subsection{Case Studies}
\label{sec:cases}
We present four representative use cases. The first two use cases are multi-application execution workflows, and the next two use cases are concurrent workloads, where multiple workloads are independent and have no dependency.

\begin{figure}[t]
    \centering
    \includegraphics[width=0.98\linewidth]{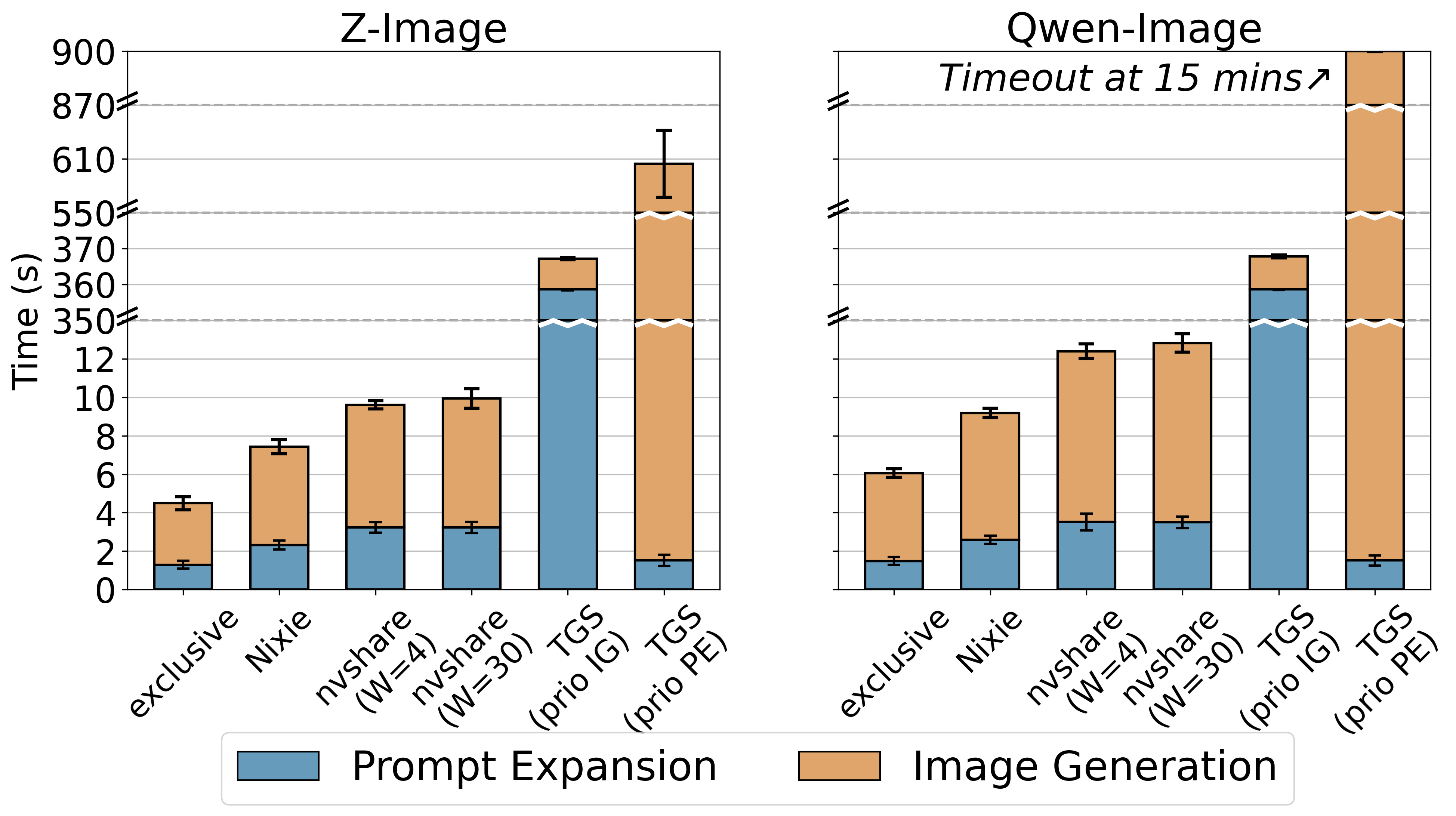}
    \vspace{-3mm}
    \caption{Using Qwen3-MoE 30B-Q6 to extend the text-to-image prompt for better image quality. Error bars represent standard deviations.}
    \label{fig:prompt_extension}
    \vspace{-5mm}
\end{figure}

\paragraph{Case \#1: Prompt-expanded image generation.}
Prompt expansion~\cite{lian2024llmgroundeddiffusionenhancingprompt, datta-etal-2024-prompt} is a standard technique for image generation. When the user wants to generate an image given a prompt, an LLM is then used to expand the prompt to make the description of the image more precise and contain more concrete details. This workflow requires a cross-application collaboration between an LLM inference application and an image generation application. 

We evaluate this workflow with \texttt{llama.cpp} and ComfyUI. For prompt expansion, we used Qwen3-MoE 30B-Q6~\cite{qwen3}. For image generation, we select two commonly used models: (1) 6B Z-Image~\cite{zimage} with BF16, and (2) 20B Qwen-Image~\cite{qwenimage} with FP8. We use the DreamBench++ dataset~\cite{dreambench} for our prompts. We compare \sys against nvshare under two window configurations: W = 4s, which matches \sys's preemption threshold of the highest priority, and W = 30s, which is the default setting of nvshare. Since TGS requires explicit priority assignment, we include two variants: one that prioritizes image generation (prio IG) and one that prioritizes the LLM-based prompt expansion (prio PE).

\autoref{fig:prompt_extension} reports the results. \sys is $1.4\times$ and $1.3\times$ faster than nvshare (W = 4) and achieves 65.9\% and 60.4\% of the performance of using two dedicated GPUs for Qwen-Image and Z-Image, respectively. Notably, even though ComfyUI frequently allocates and frees GPU memory during Qwen-Image generation, \sys maintains higher performance, indicating that the performance of \sys is robust for memory allocation and deallocation on the data path. 

TGS exhibits significantly degraded performance and cannot finish image generation for Qwen-Image after 15 minutes. There are two primary reasons: (1) TGS assumes that the high-priority job must sustain a constant throughput, which does not hold for most consumer-oriented applications. Consequently, the TGS rate limiter fails to function effectively in this scenario. (2) Under memory oversubscription, TGS forces low-priority jobs to access host memory directly via DMA without migration, essentially stalling computation.

\begin{figure}
\centering
\begin{subfigure}[b]{0.49\linewidth}
        \centering
        \includegraphics[width=\textwidth]{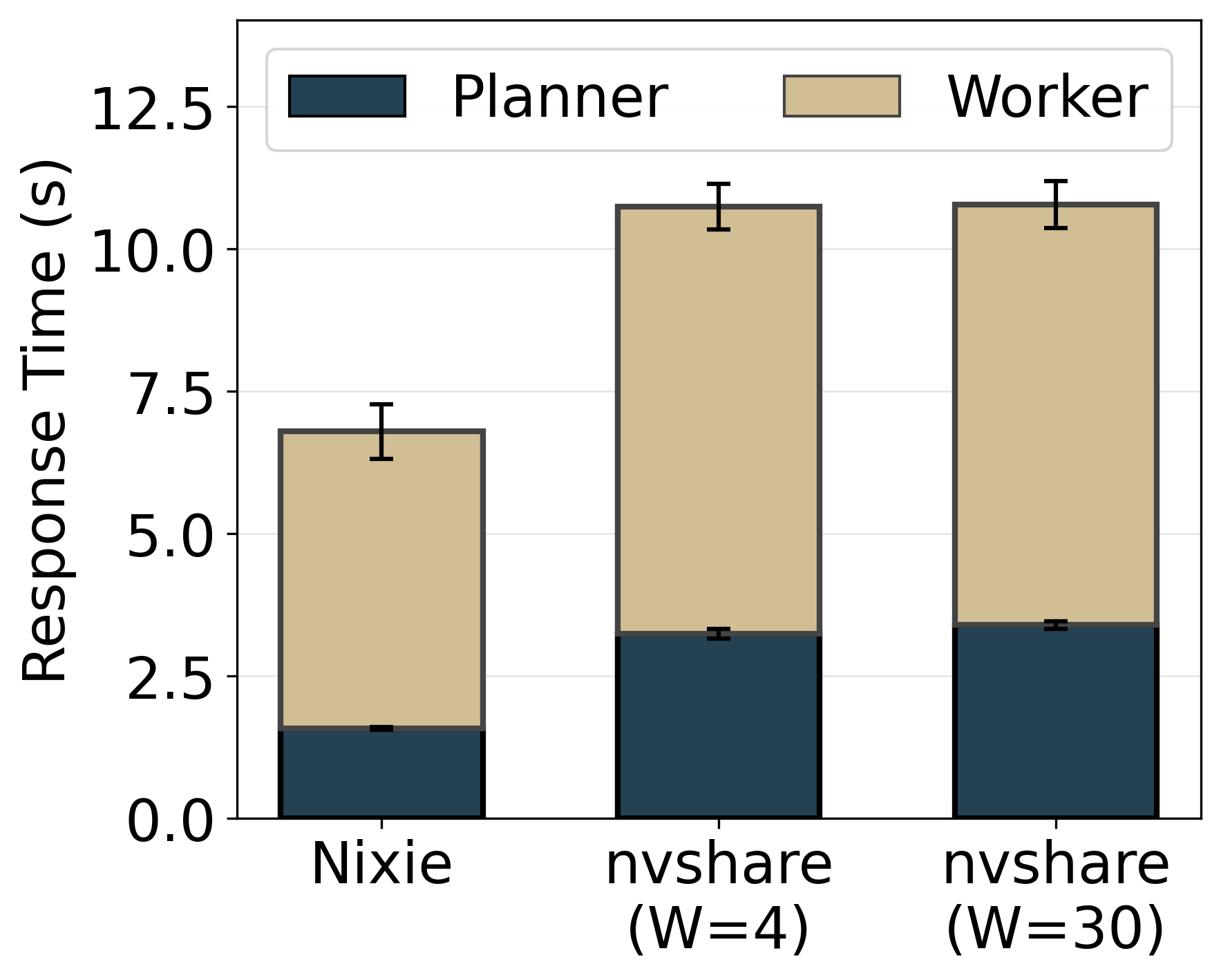}
        \caption{Overall Latency}
        \label{fig:mix_kvcomm_overall}
    \end{subfigure}
    \hfill
    \begin{subfigure}[b]{0.49\linewidth}
        \centering
        \includegraphics[width=\textwidth]{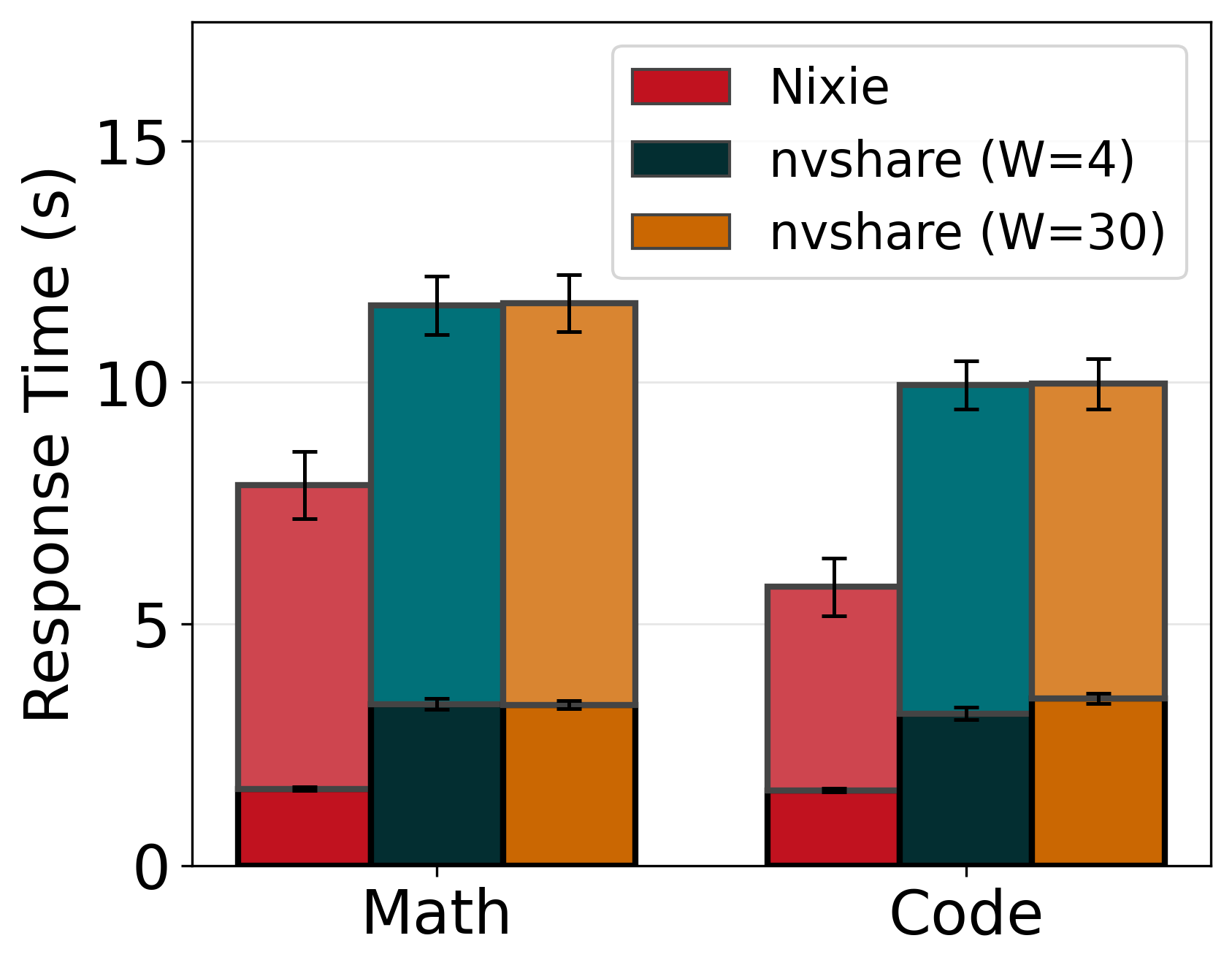}
        \caption{Latency by Request Type}
        \vspace{1.3mm}
        \label{fig:mix_kvcomm_bytype}
    \end{subfigure}
\caption{Multi-agent orchestration for math and code tasks. Error bars represent standard deviations.}
\label{fig:model_orch}
\vspace{-5mm}
\end{figure}

\paragraph{Case \#2: Multi-agent systems.}
Multi-agent systems are an increasingly common way to structure complex LLM applications.  
Instead of relying on a single monolithic model, a \emph{planner} agent decomposes a user request into subtasks and dispatches each subtask to a specialized \emph{worker} model.
This planner-worker orchestration pattern appears in current coding assistants, tool-using agents and many recent multi-LLM workflows, and serves as a canonical multi-agent workload\cite{openagi, megaagent, metagpt}.

We instantiate this model-orchestration workload using the open-source multi-agent framework KVCOMM~\cite{kvcomm}. For routing, we use Qwen3 14B-Q8 to decide the best downstream model.
Following model-level benchmarks~\cite{gemma3,qwen3,qwen3vl,qwen3coder}, we select Qwen3-VL 32B-Q6 as the math expert and Qwen3-Coder 30B-Q6 as the coding expert.
All models are served through \texttt{llama.cpp} on a single GPU; the only difference between configurations is the GPU sharing mechanism. We compare \sys against nvshare with W=4 and W=30.
We issue a mixed stream of math and coding requests (shuffled from GSM8K~\cite{gsm8k} and HumanEval-FIM~\cite{ humaneval-fim}) and record end-to-end latency per request, as well as its breakdown into planner and worker components.

\autoref{fig:model_orch} shows our results.
Across all requests, \sys is 1.6x faster than both nvshare configurations. The gains are consistent across request types as well: for math and coding tasks, \sys is $1.5\times$ and $1.7\times$ faster, respectively. These results indicate that \sys offers strong flexibility across multiple applications, enabling efficient model orchestration on a single shared GPU without sacrificing the interactivity of the multi-agent workflow.

\begin{figure}
\centering
\includegraphics[width=0.98\linewidth]{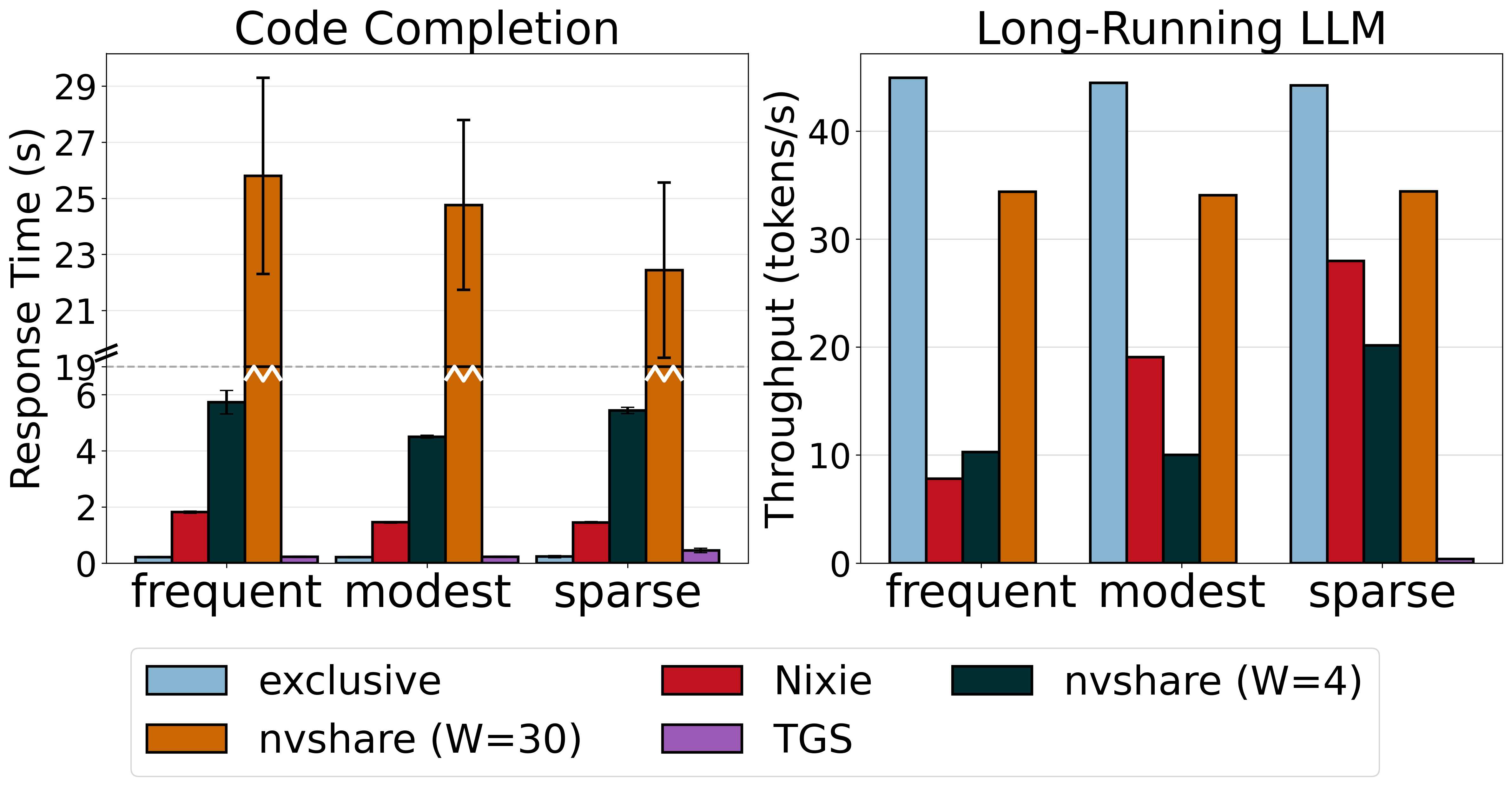}
\vspace{-3mm}
\caption{Code completion with long-running generation. Error bars represent standard deviations.}
\label{fig:coding}
\vspace{-3mm}
\end{figure}

\paragraph{Case \#3: Code completion with long-running job.}

LLM-assisted code completion is a representive latency-sensitive task requiring immediate responsiveness. We expect \sys to naturally prioritize it over long-running jobs, such as autonomous agent operations or extended video generation tasks. To evaluate \sys's prioritization behavior, we run code completion with a long-running agentic workload. We use Qwen3-Coder 30B-Q6 for Fill-In-Middle (FIM) completion and Gemma3 27B-Q8 to emulate an LLM-based agent. HumanEval-FIM~\cite{humaneval-fim} is used as the input for code completion. Since the frequency and interval between code completion requests vary depending on individual developers and their tooling implementations, we create three scenarios to represent different usage patterns: frequent (1-second intervals), modest (3-second intervals), and sparse (6-second intervals).

Figure \ref{fig:coding} shows the response time of code completion and the throughput of the long-running model. \sys achieves an average response time of 1.8s for frequent completion scenario and 1.4s for modest and sparse scenarios. In contrast, nvshare (W=4) requires 4.5s to 5.7s to complete an interactive request depending on the frequency. \sys is $3.1\times$ to $3.8\times$ compared to nvshare (W=4). When nvshare uses the default W=30, the latency exceeds 20s, rendering LLM code completion impractical for real-world use cases.

Regarding the long-running job throughput, \sys is 90.6\% and 39.5\% higher than nvshare (W=4) in the modest and sparse scenarios respectively. In the frequent scenario, \sys throughput is 23.5\% lower, which is expected: \sys prioritizes the interactive jobs and allocate more GPU time for code completion.
TGS achieves similar performance for code completion task, but fails to allow the long-running model to run. TGS essentially degrades to running only the high priority application.

\begin{figure}
\centering
\includegraphics[width=0.95\linewidth]{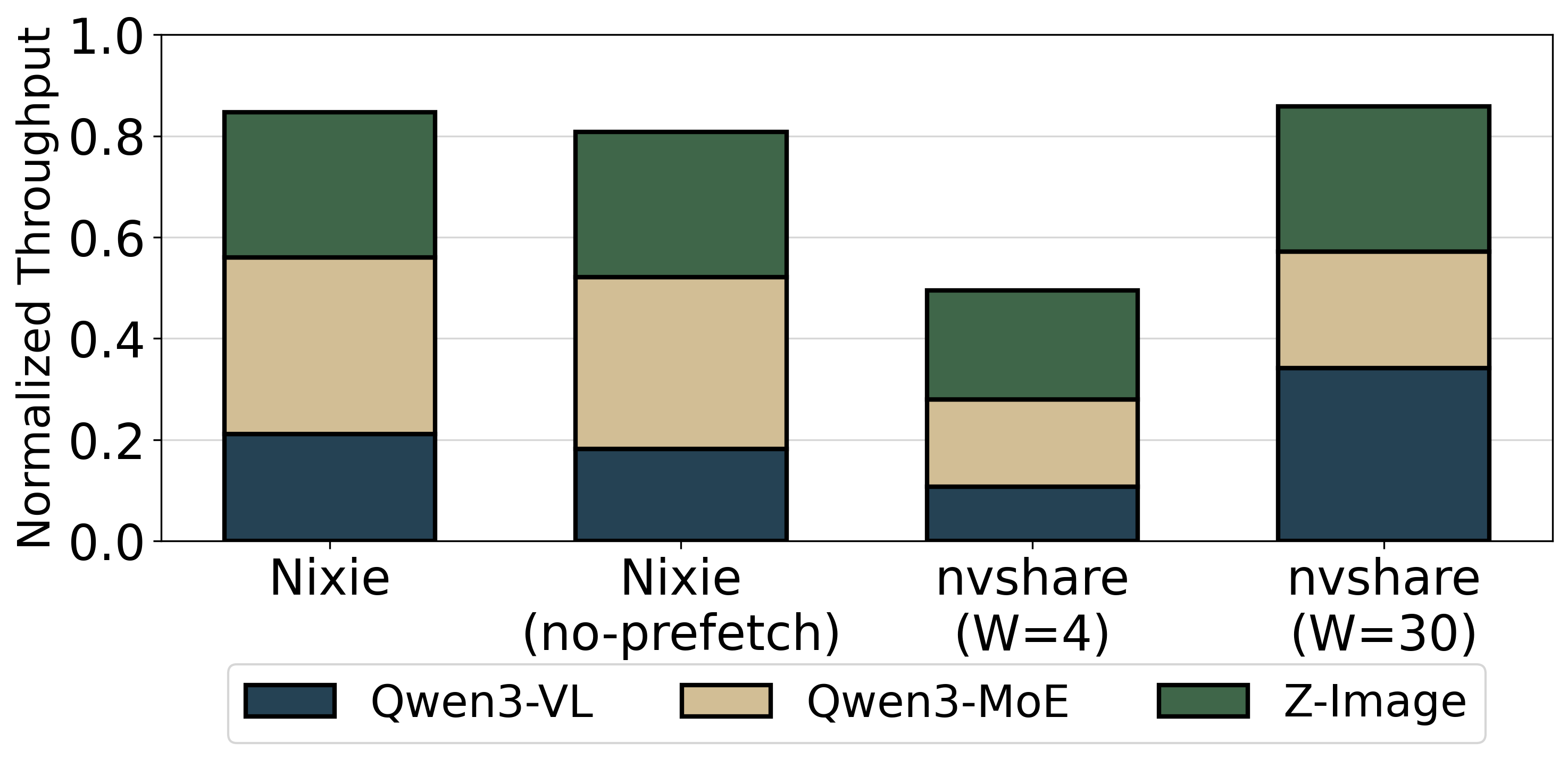}
\vspace{-3mm}
\caption{Different Batching Jobs}
\label{fig:batching}
\vspace{-5mm}
\end{figure}

\paragraph{Case \#4: Many Batching Tasks.}
While \sys targets interactive workloads, some GPU jobs are sufficiently heavy that they must run for extended periods (e.g., long-document processing, high-volume generation). We therefore evaluate both throughput and fairness in a setting where multiple batched jobs execute concurrently. 

We run three jobs: Qwen3-MoE 30B-Q6, Qwen3-VL 32B-Q6 (long-running LLM request with batch size 1), and Z-image (batch size 8). We also ablate the effectiveness of our auto-prefetching mechanism in the scheduler. We compare our solutions with nvshare configured with W=4 and W=30. All experiments are conducted over a 300-second duration. To account for the heterogeneity of metrics across different applications, we additionally run each application in isolation as a reference, and report normalized throughput relative to its profiled standalone performance.

\autoref{fig:batching} shows the result. \sys achieves throughput comparable to nvshare (W=30) achieving 85\% of the ideal throughput, even when \sys employs an adaptive policy to prioritize responsiveness, whereas nvshare utilizes a fixed long time quantum for throughput. Fairness is effectively maintained across all tested configurations. Auto-prefetching yields an additional 5\% throughput improvement over the no-prefetch variant, despite context switches already being infrequent in this scenario. In contrast, nvshare (W=4) only realizes 49.5\% throughput because of the overhead brought by frequent context switches.

\subsection{Performance on Other Hardware}

We evaluate \sys on another server equipped with two AMD EPYC 7352 24-Core Processor CPUs, 1TB of 16-channel DDR4 memory at 3200MT/s, and 8x RTX A5000 GPU (24GB) connected via PCIe 4.0x16. It runs Ubuntu 22.04 with CUDA 12.4 and Nvidia driver version 550.67. 

On this platform, we reuse the multi-agent orchestration workload from \textbf{Case \#2}. 
Given RTX A5000 has smaller memory, we use Qwen3 14B-Q4 as the planner, Qwen3-VL 32B-Q4 as the math worker, and Qwen3-Coder 30B-Q4 as the coding worker. We add \emph{exclusive} variant here given more available GPUs. \sys shows $3.4\times$ faster than nvshare in this case and achieves 73\% performance of \emph{exclusive}. UVM-based solution slows down significantly for this setup. We observed increase in PCIe idleness on this hardware. We suspect the root cause is the older NVIDIA driver, and the correspondingly older UVM implementation.

\begin{figure}
\centering
\begin{subfigure}[b]{0.49\linewidth}
        \centering
        \includegraphics[width=\textwidth]{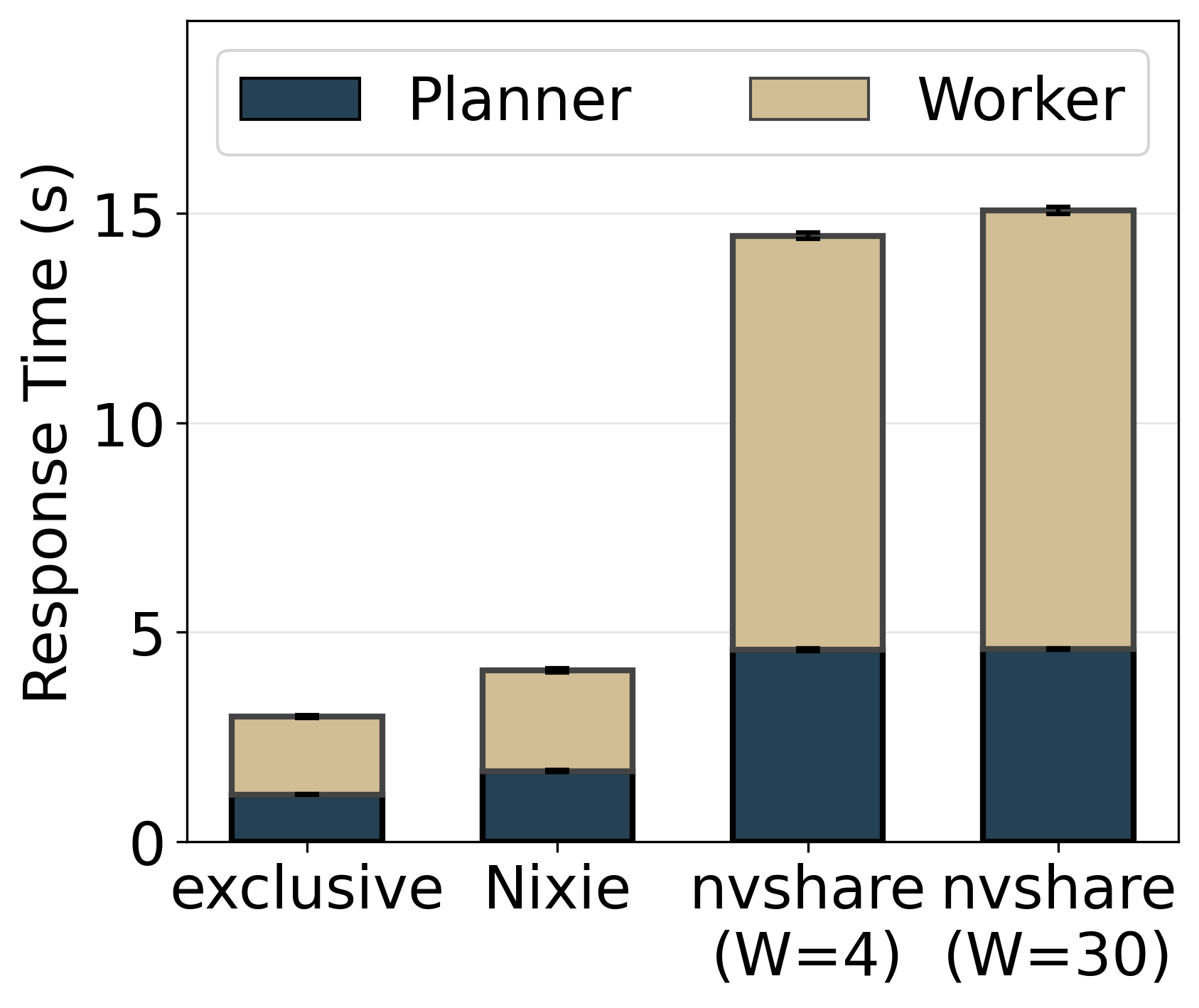}
        \caption{Overall Latency}
        \label{fig:mix_kvcomm_a5000_overall}
    \end{subfigure}
    \hfill
    \begin{subfigure}[b]{0.49\linewidth}
        \centering
        \includegraphics[width=\textwidth]{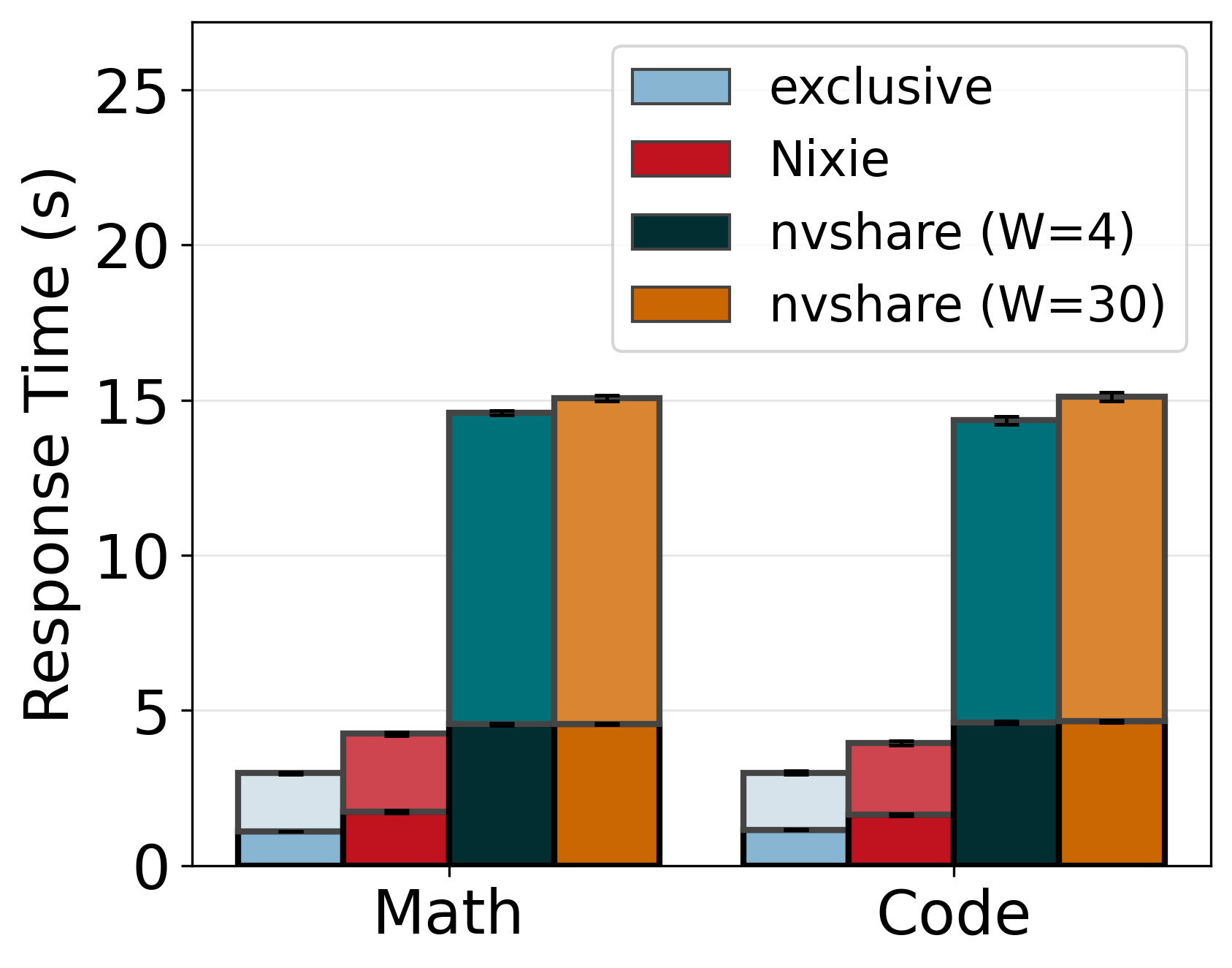}
        \caption{Latency by Request Type}
        \vspace{1.3mm}
        \label{fig:mix_kvcomm_a5000_bytype}
    \end{subfigure}
\caption{Case \#2 on RTX A5000 testbed. Error bars represent standard deviations.}
\label{fig:model_orch_a5000}
\vspace{-5mm}
\end{figure}

\section{Discussion}

\paragraph{Generalizability to other operating systems.}
\sys leverages CUDA VMM API, which is available on both Linux and Windows. We have prototyped \sys on Linux, and we believe that it is feasible to port \sys to Windows, which remains the most widely used operating system for consumers. In contrast, UVM does not support memory oversubscription on Windows\cite{uvm_windows}, and adding such support would require substantial engineering effort in the GPU driver stack in Windows. This distinction makes \sys a more practical solution for consumer platforms. Furthermore, similar VMM APIs also exist for AMD GPUs~\cite{amd_hip_virtual_memory}.

\paragraph{White-Box solutions.}
\sys is a fully transparent solution that operates without any awareness of tensor semantics. In principle, certain immutable data (e.g., model weights) could be freed directly on the GPU without migrating it back to CPU memory, provided the application retains a separate CPU-side copy. For example, CUDA memory copies inherently leave duplicate data in both CPU and GPU memory. We did not pursue this direction because it would require semantic knowledge of application data structures, compromising transparency. However, 
incorporating such semantic hints represents an interesting direction for future work.

\paragraph{Co-locating with small models.}
\sys's target workloads is where each application utilizes most of the GPU memory, so temporal multiplexing is the only approach. If we have a set of small models, then it is possible to do spatial multiplexing and having different models execute at the same time like existing works in datacenter GPU multiplexing~\cite{nexus, shepherd, clockwork}. How to integrate spatial multiplexing into \sys is a promising future direction.

\paragraph{Security.}
\daemon shares the same UID with applications, and applications share the same region of CPU pinned memory for storing migrated GPU data. A malicious process may access the shared data region inside its own address space. However, Linux uses UID for shared memory access control, which means any process under the same UID can access CPU pinned memory, whether they are shared by all processes or each process has its dedicated CPU pinned memory. Since \sys is for a single-user environment rather than multi-tenancy, this is outside our threat model.

\section{Related Works}

\paragraph{GPU multiplexing before the era of LLMs.}
GPU multiplexing has been extensively studied prior to the rise of LLMs. For traditional computer vision workloads, multiplexing allows significant consolidation of hardware by serving many models on shared GPUs (e.g., Nexus~\cite{nexus}, Shepherd~\cite{shepherd}, Clockwork~\cite{clockwork}). In these systems, GPU memory stores multiple models simultaneously, and incoming requests are batched and executed across one or more models. Reducing the number of co-located models or lowering batch sizes helps alleviate GPU memory pressure. In this setting, the dominant challenge is maximizing GPU compute efficiency.
In contrast, \sys targets temporal, not spatial, multiplexing: rather than having the GPU inference many models, \sys time-shares it across large, memory-intensive applications whose working sets approach the GPU’s full capacity.

\paragraph{GPU multiplexing in the era of LLMs and other large foundation models.}
LLMs and other large foundation models have fundamentally reshaped the GPU multiplexing landscape. Attention computation is highly memory-intensive, especially with long contexts. Even a single LLM inference often consumes nearly all available GPU memory. As a result, recent research has pivoted toward temporal multiplexing and techniques for handling GPU memory oversubscription.

PipeSwitch~\cite{pipeswitch} introduces a model-aware, layer-based task-switching mechanism, but it requires application modifications. Further, it is less effective as the ratio of migration time to model execution time grows in modern foundation models. ServerlessLLM~\cite{serverlessllm}, Aegaeon~\cite{aegaeon}, and Prism~\cite{prism} address oversubscription by offloading less frequently accessed model components (e.g., entire models or KV caches) to CPU memory or disk, primarily to meet provider's Service Level Objectives.
These systems are designed for datacenter GPUs, where workloads are controlled by a single provider and can be integrated directly into the training and inference stack. However, they do not translate to consumer GPUs, where workloads are diverse, applications are not centrally managed, and users expect transparent solutions.

\paragraph{Improving UVM.}
UVM provides transparent abstraction for GPU memory oversubscription, which is appealing.
Systems such as DeepUM~\cite{deepum} optimize UVM performance by incorporating tensor-level prefetching. G10~\cite{g10} adds disks to UVM and uses compiler to help manage pipelined data transfer. These solutions require hardware changes or driver changes to GPUs. Further, \sys is no a purely a virtual memory system. It does not depend on page faults and can coordinate compute and memory management at the same time to mitigate memory thrashing.

\paragraph{GPU scheduling.}
Multiple works discussed multitasking scheduling. XSched~\cite{xsched} presents a framework that allows for preemptive XPU scheduling. Autellix~\cite{autellix} incorporates application dependencies into LLM scheduling. Our scheduling policy is quite simple, and we do not need any application-level information. Our scheduling goal is to prioritize interactive applications.

\section{Conclusion}
Consumer machines increasingly execute large ML workloads such as LLM inference, text-to-image generation, and interactive image editing. Unlike datacenter environments, consumer GPUs must handle multiple applications concurrently, where each application often nearly saturate GPU memory.
We have built \sys, enabling efficient and fully transparent temporal multiplexing on consumer GPUs without requiring any modifications to applications or GPU drivers. \sys coordinates GPU memory allocation with kernel-launch behavior to minimize thrashing. \sys's scheduler further enhances interactivity by automatically identifying and prioritizing latency-sensitive applications.
\sys improves latency for interactive code-completion workloads by up to 3.8$\times$, while reducing CPU pinned-memory usage by up to 66.8\% under equivalent latency constraints. Our source code will be publicly available.

\newpage
\bibliographystyle{plain}
\bibliography{ref}

\end{document}